\documentclass[structabstract]{aa}
\usepackage{graphicx}
\usepackage{natbib}
\usepackage[dvipsnames]{xcolor}
\usepackage{hyperref}
\usepackage{txfonts}
\usepackage{longtable}

\title{Discovery of an asteroid family linked to (22) Kalliope\\ and its moon Linus}

\titlerunning{Discovery of an asteroid family linked to (22) Kalliope and its moon Linus}

\author{
    M.~Bro\v{z}\inst{\ref{prague}}         \and % multipole model
    M.~Ferrais\inst{\ref{lam}}             \and % LP
    P.~Vernazza\inst{\ref{lam}}            \and % LP
    P.~\v{S}eve\v{c}ek\inst{\ref{prague}}  \and % Opensph
    M.~Jutzi\inst{\ref{bern}}                   % discussion of shapes
}

\institute{
     Charles University, Faculty of Mathematics and Physics, Institute of Astronomy, V~Hole{\v s}ovi{\v c}k{\'a}ch 2, 18000 Prague, Czech Republic%
     \label{prague}%
     \and %---- Vernazza, Jorda, Ferrais, Fusco, Fetick, Drouard
     Aix Marseille Univ, CNRS, LAM, Laboratoire d'Astrophysique de Marseille, Marseille, France
     \label{lam}
     \and %---- Jutzi
     University of Bern, Physics Institute, NCCR PlanetS, Gesellschaftsstrasse 6, 3012, Bern, Switzerland
     \label{bern}
}

\date{Received x-x-2022 / Accepted x-x-2022}

\abstract
   {}
   {
According to adaptive-optics observations by Ferrais et al.,
(22) Kalliope is a 150-km, dense and differentiated body.
Here, we interpret (22) Kalliope
in the context of bodies in its surroundings.
While there is a known moon Linus,
with a 5:1 size ratio,
no family has been reported in the literature,
which is in contradiction with the existence of the moon.
   } 
   {
Using the hierarchical clustering method (HCM)
along with physical data,
we identified the Kalliope family.
Previously, it was associated to (7481) San Marcello.
We then used various models (N-body, Monte-Carlo, SPH) of its
orbital and collisional evolution,
including the break-up of the parent body,
to estimate the dynamical age of the family
and address its link to Linus.
   }
   {
The best-fit age is $(900\pm100)\,{\rm My}$ according to our collisional model,
in agreement with the position of (22) Kalliope,
which was modified by chaotic diffusion due to $4{-}1{-}1$ three-body resonance
with Jupiter and Saturn.
It seems possible to create Linus and the Kalliope family at the same time,
although our SPH simulations show a variety of outcomes,
for both satellite size and the family size-frequency distribution.
The shape of (22) Kalliope itself was most likely affected
by gravitational reaccumulation of `streams',
which creates characteristic hills observed on the surface.
If the body was differentiated, its internal structure is surely asymmetric.
   }
   {}
 

\begin{document}

\maketitle

%%%%%%%%%%%%%%%%%%%%%%%%%%%%%%%%%%%%%%%%%%%%%%%%%%%%%%%%%%%%%%%%%%%%%%%%

\section{Introduction}

(22) Kalliope is the second largest M-type asteroid in the main belt,
after (16) Psyche, and as such, it has been a promising target for spatially resolved
observations \citep{Sokova_2014Icar..236..157S,Drummond_2021Icar..35814275D}.
Recent VLT/SPHERE adaptive-optics observations of Kalliope by \cite{Ferrais_2022}, 
along with archival astrometry and interferometry of its moon Linus,
led to precise estimates of its fundamental physical properties
(size, shape, volume, mass, density).
Its exceptional density ($\geq4\,{\rm g}\,{\rm cm}^{-3}$),
the highest found so far among asteroids \citep{Vernazza_2021A&A...654A..56V},
in tandem with its low radar albedo ($0.18\pm 0.05$; \citealt{Shepard2015}),
corresponding to a metal-poor (silicate-rich) surface,
strongly suggests a differentiated interior.
Regarding the nature of the silicates present at the surface, near-infrared spectroscopic observations suggest that these may comprise low-calcium pyroxene and possibly hydrated silicates \citep{Hardersen2011, Usui2019}.
Overall, Kalliope's case may be similar to that of Mercury,
for which the high density is explained by a giant collision and mantle stripping
\citep{Asphaug_Reufer_2014NatGe...7..564A},
although alternative explanations exist
(e.g., \citealt{Broz_2021NatAs...5..898B}).

The Kalliope--Linus binary system is also exceptional,
especially because Linus is by far the largest asteroid moon
\citep{Descamps_2008Icar..196..578D},
possessing a diameter $(28\pm2)\,{\rm km}$ and primary/secondary ratio of approximately 5:1.
In this sense, it is similar to the Earth--Moon system (cf.~4:1).
According to the accepted rule, `every giant moon requires a giant impact'
\citep{Hartmann_1975Icar...24..504H,Durda_2004Icar..170..243D}.
For asteroids located on stable orbits within the main belt,
it inevitably implies, `every giant moon requires an asteroid family',
because fragments ejected during break-up often land on stable orbits.
This was our motivation to search for as-of-yet unknown family
in the vicinity of (22) Kalliope.

%%%%%%%%%%%%%%%%%%%%%%%%%%%%%%%%%%%%%%%%%%%%%%%%%%%%%%%%%%%%%%%%%%%%%%%%

\section{Observed Kalliope family}\label{observed}

(22) Kalliope (osculating $a = 2.909\,{\rm au}$, $e = 0.098$, $I = 13.701^\circ$)
is located in the so called `pristine zone' \citep{Broz_2013A&A...551A.117B}
of the main belt,
which is surrounded by strong mean-motion resonances with Jupiter, namely by
the 5:2 resonance at 2.82\,au, and
the 7:3 resonance at 2.96\,au.
Consequently, it is strongly depleted, because small km-sized asteroids
drifting by the Yarkovsky effect have relatively shorter lifetimes,
compared to the middle and outer belts.

\begin{figure}
\centering
\includegraphics[width=9cm]{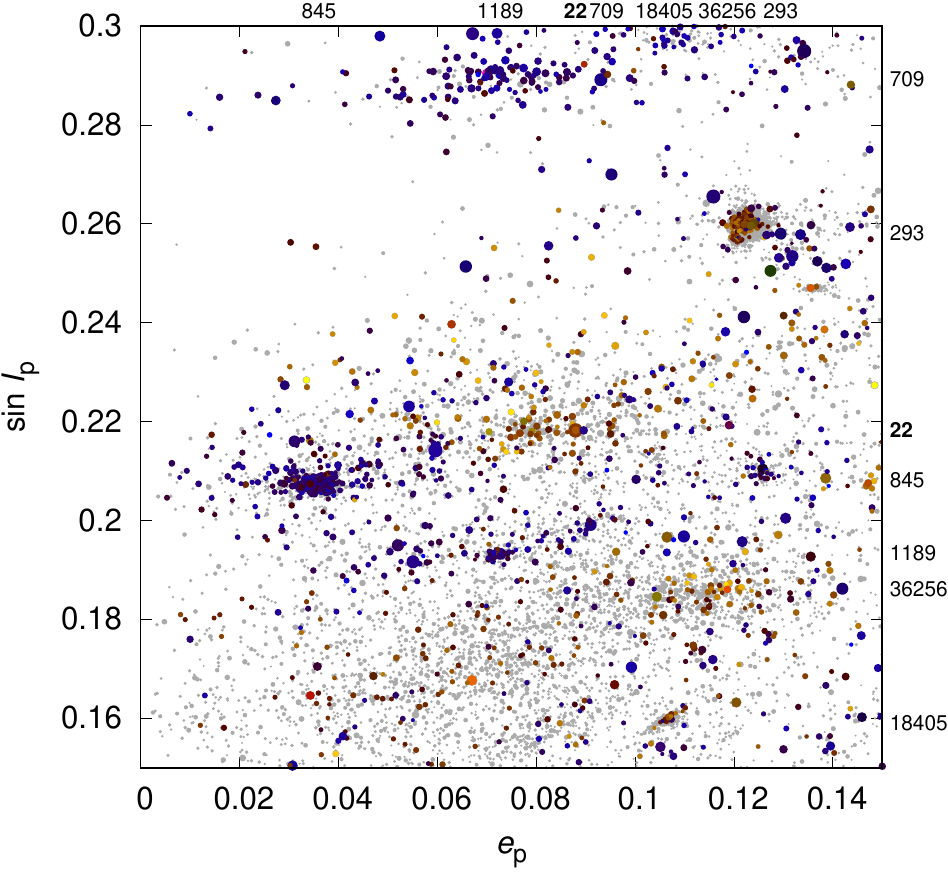}
\includegraphics[width=9cm]{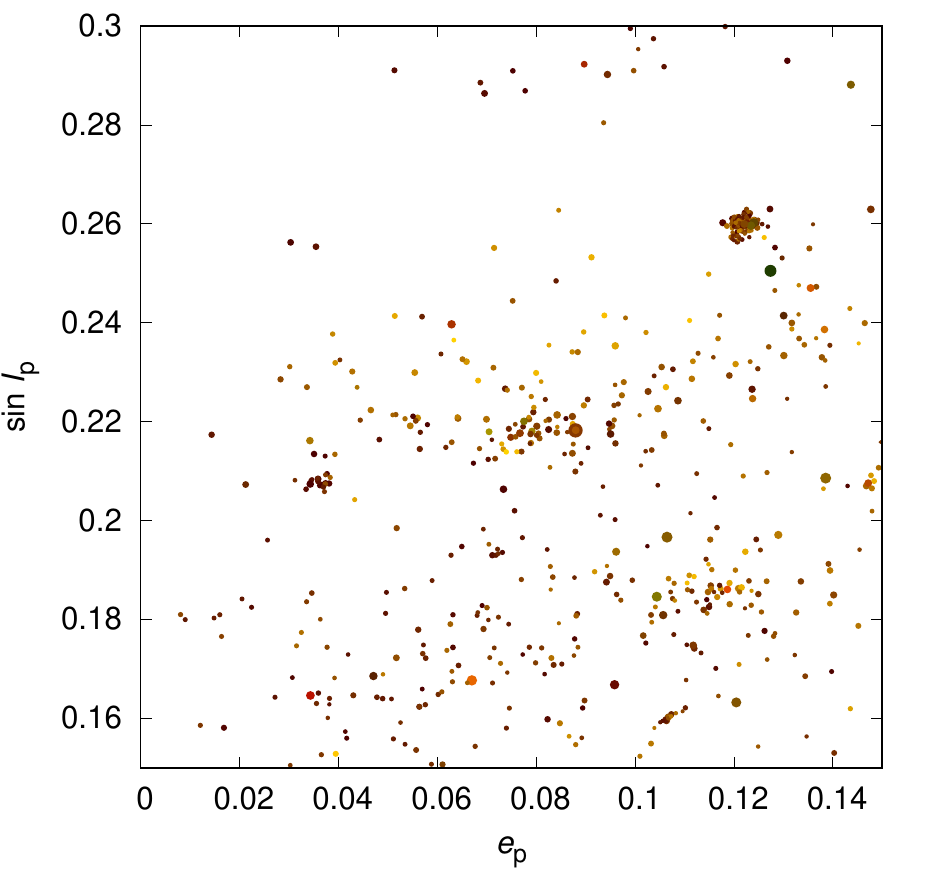}
\caption{
Observed proper eccentricity $e_{\rm p}$ vs. sine of proper inclination $\sin I_{\rm p}$
of bodies in the `pristine zone';
with the proper semimajor axis $a_{\rm p} \in (2.82; 2.96)\,{\rm au}$.
All bodies are plotted (top) and a subset having the geometric albedo
$p_V \in (0.1; 0.35)$ (bottom),
according to the WISE catalogue \citep{Nugent_2015ApJ...814..117N}.
Colours correspond to $p_V$
(blue$\,\rightarrow\,$yellow).
If $p_V$ is unknown, the colour is gray.
The high-albedo family previously designated (7481) San~Marcello = FIN~626
is now associated to (22)~Kalliope (bold number).
Numerous known families are indicated (numbers at the border),
namely (293), (709), (845), (1189), (18405), and (36256).
}
\label{ei_wise}
\end{figure}

We used recent catalogues of proper elements
\citep{Knezevic_Milani_2003A&A...403.1165K,Radovic_2017MNRAS.470..576R,Novakovic_2019EPSC...13.1671N}
and of albedos
\citep{Nugent_2015ApJ...814..117N,Usui_2011PASJ...63.1117U}
to plot Fig.~\ref{ei_wise}.
One can immediately identify a number of known families
\citep{Nesvorny_2015aste.book..297N},
including the one denoted (7481) San~Marcello,
which is surprisingly close to (22) Kalliope.
For reference, its family identification number (FIN) is 626.

\begin{figure}
\centering
\includegraphics[width=9cm]{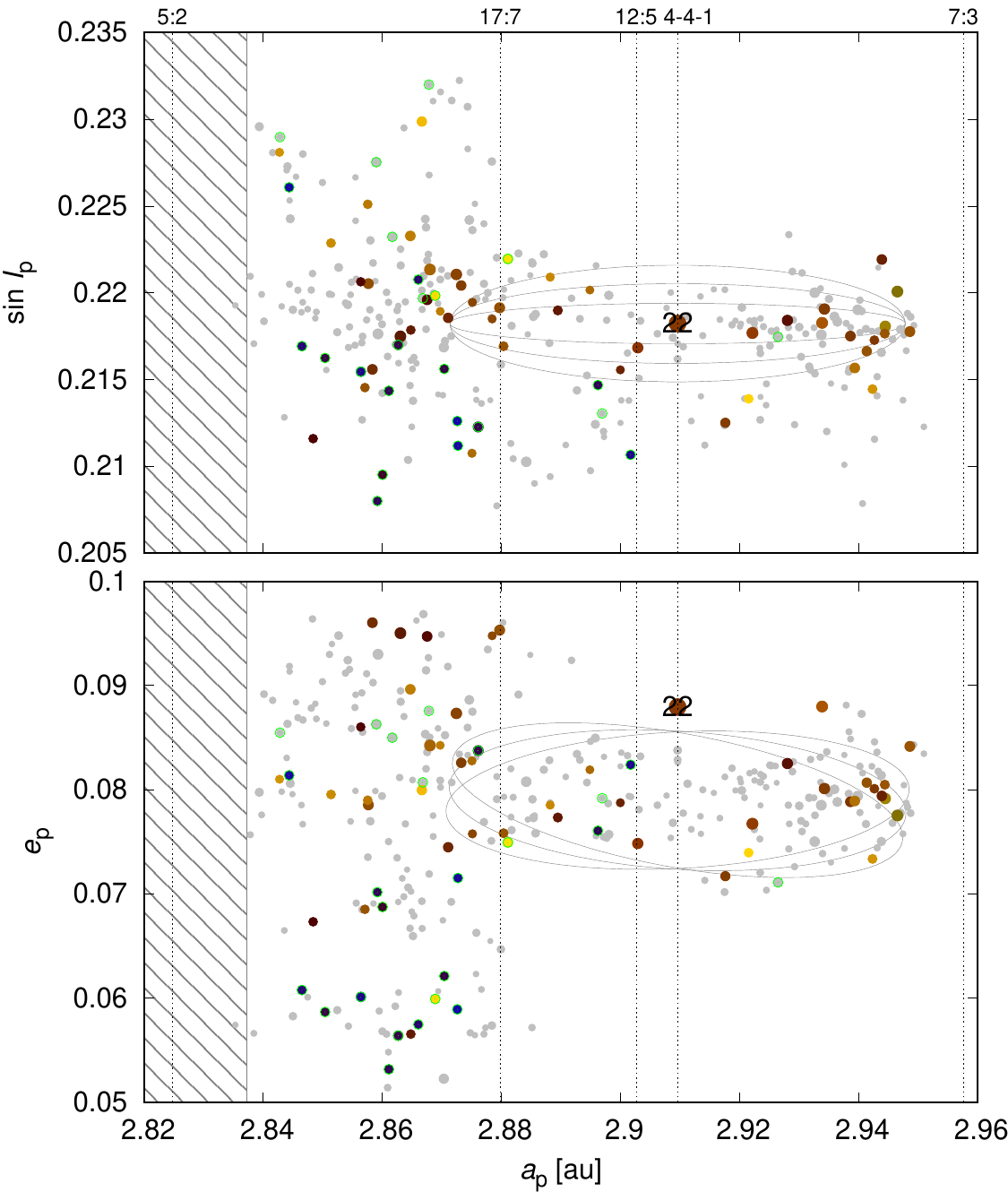}
\caption{
The observed Kalliope family in the space of proper elements
$a_{\rm p}$, $e_{\rm p}$ (bottom), and $\sin I_{\rm p}$ (top).
It was identified by the hierarchical clustering,
for the cutoff velocity $v_{\rm cut} = 75\,{\rm m}\,{\rm s}^{-1}$.
Interlopers are indicated by green circles.
The mean-motion resonances 5:2, 17:7, 12:5, $4{-}4{-}1$, and 7:3
are plotted by dotted lines.
The extent of the 5:2 is hatched.
(22)~Kalliope is located in the three-body resonance \hbox{$4{-}4{-}1$}
with Jupiter and Saturn, which explains why it is separated from the family.
The iso-velocity ellipses were computed for the escape velocity
$v_{\rm esc} = 116\,{\rm m}\,{\rm s}^{-1}$
and specific values of the true anomaly and the argument of pericentre:
$f = 90^\circ, 100^\circ, 110^\circ$,
$\omega+f = 60^\circ, 70^\circ, 80^\circ$.
}
\label{aei2_vgauss}
\end{figure}

Moreover, we realized that the semimajor axis of (22) Kalliope
coincides with the three-body resonance $4{-}4{-}1$
with Jupiter and Saturn at 2.91\,au
\citep{Nesvorny_1998AJ....116.3029N},
and it is shifted in eccentricity by 0.01
with respect to the family FIN~626 (cf.~Fig.~\ref{aei2_vgauss}).
This is most likely due to chaotic diffusion,
(see confirmation in Sec.~\ref{orbital}).
In this work, we thus suggest that the whole family 626
should be associated to (22) Kalliope
(FIN always remains the same).

We used the hierarchical clustering method (HCM; \citealt{Zappala_1995Icar..116..291Z})
to extract the family.
However, our initial body was still (7481), not (22),
because it is too separated.
The maximum possible cutoff velocity was 
$v_{\rm cut} = 75\,{\rm m}\,{\rm s}^{-1}$;
(22) was added `manually'.
Interlopers were removed automatically,
if they did not fulfil our criteria for members:
the visible albedo $p_V \in (0.1; 0.35)$
the colour index $a^\star \in (-0.5; 0.025)\,{\rm mag}$.
The result is shown in Fig.~\ref{aei2_vgauss}.
The overall extent roughly corresponds to the escape speed from the parent body, i.e.,
$v_{\rm esc} = 116\,{\rm m}\,{\rm s}^{-1}$.
The Kalliope family exhibits a typical `V-shape',
\citep{Vokrouhlicky_2006Icar..182..118V},
with the centre $a_{\rm c} = 2.9095\,{\rm au}$
and the parameter $C = 1.5\cdot 10^{-4}$.
(22) Kalliope is located close to this centre,
because a semimajor axis is not changed by chaotic diffusion
in a mean motion resonance.
Given the median albedo $p_V = 0.195$
and the bulk density $\rho = 4.1\,{\rm g}\,{\rm cm}^{-3}$
(according to the ADAM shape),
the {\em upper limit\/} for the age is \citep{Nesvorny_2015aste.book..297N}:
\begin{equation}
t_{\uparrow} = 1\,{\rm Gy} {C\over 10^{-4}} \left({a_{\rm c}\over 2.5\,{\rm au}}\right)^2 {\rho\over 2.5\,{\rm g}\,{\rm cm}^{-3}} \left({0.2\over p_V}\right)^{1/2} = 3.4\,{\rm Gy}\,.
\end{equation}
A larger dispersion in eccentricity (0.03 vs. 0.01) is observed
on the left-hand side of the 17:7 mean-motion resonance with Jupiter,
presumably due to the Yarkovsky drift across the resonance.
Consequently, majority of family members were originally located
on the right-hand side.

\begin{figure*}
\centering
\begin{tabular}{c@{}c@{}c}
\kern.7cm continuous &
\kern.7cm with Linus &
\kern.7cm populous \\
\includegraphics[width=6.0cm]{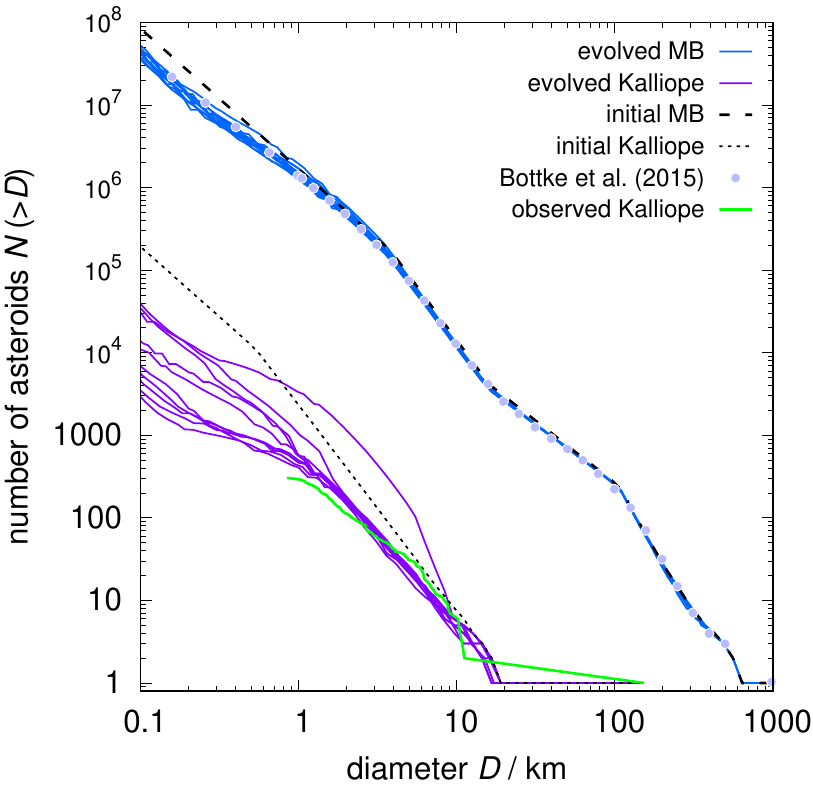} &
\includegraphics[width=6.0cm]{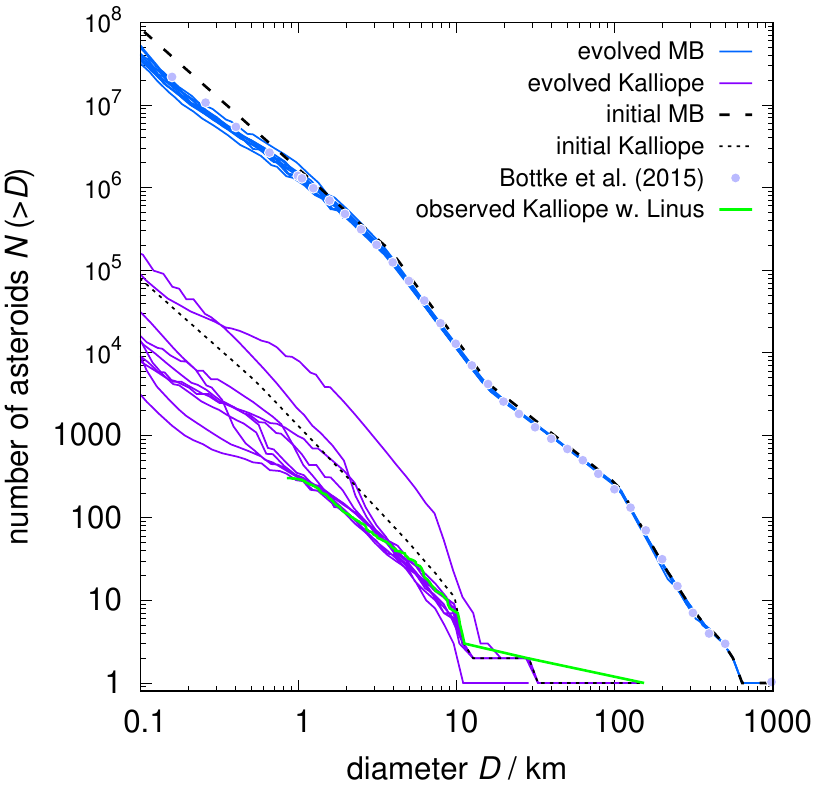} &
\includegraphics[width=6.0cm]{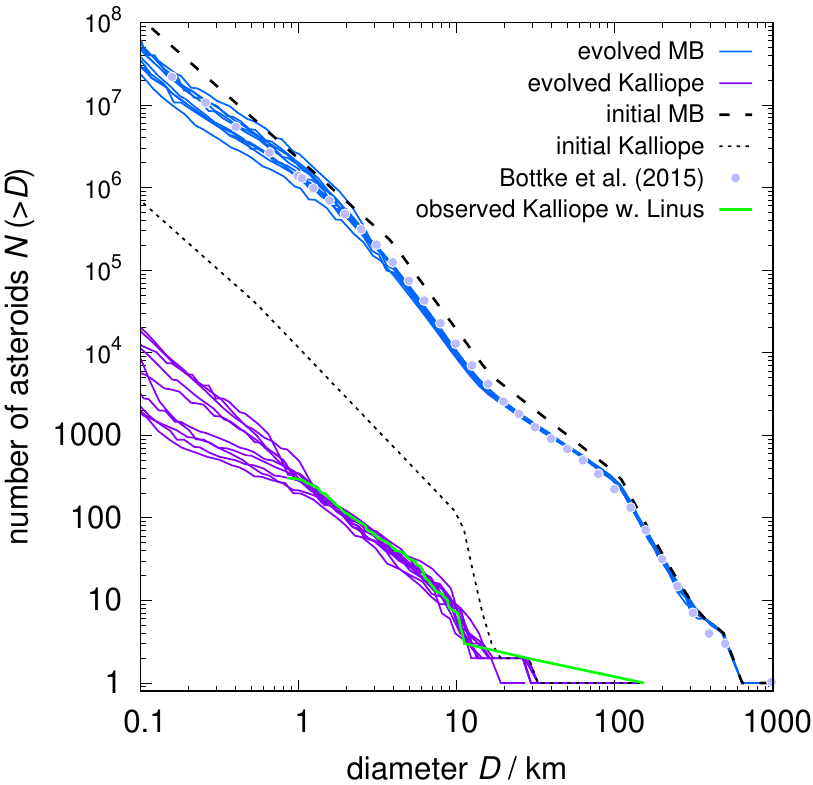} \\
\end{tabular}
\caption{
Collisional evolution of the main belt (blue) and the Kalliope family (pink).
Cumulative size-frequency distributions $N({>}D)$ are plotted.
Observed main belt population is taken from \cite{Bottke_2015aste.book..701B},
observed Kalliope family from this work (green).
We assumed three different initial conditions (black dashed):
continuous SFD (left),
with Linus and depleted $D > 10$-km bodies (middle),
populous SFD (right).
Each simulation was run 10 times (multiple pink lines), in order to account for
stochasticity of collisions.
The respective best-fit ages are
$800\,{\rm My}$, $900\,{\rm My}$, up to $3.4\,{\rm Gy}$.
}
\label{37_MB_Kalliope_LINUS_sfd_0900}
\end{figure*}

The size-frequency distribution (SFD) was computed from 302 members.
It exhibits a steep part (slope $-3.0$)
and a very shallow part ($-1.5$) below $D < 5\,{\rm km}$
(Fig.~\ref{37_MB_Kalliope_LINUS_sfd_0900}),
which is typical for dynamically depleted populations.

A preliminary comparison to a set of SPH simulations
\citep{Durda_2007Icar..186..498D} indicates
a parent body size of $D_{\rm pb} = (157\pm2)\,{\rm km}$,
the projectile size $d = (29\pm 10)\,{\rm km}$,
the largest remnant mass ratio $M_{\rm lr}/M_{\rm pb} = 0.92\pm0.05$,
the largest fragment $M_{\rm lf}/M_{\rm pb} = 0.00036\pm0.00010$, and
the specific energy $Q/Q^\star = 0.10\pm0.05$,
where $Q^\star$ corresponds to the scaling law of \cite{Benz_Asphaug_1999Icar..142....5B}.
Because the outcome depends on so many parameters,
including the impact speed~$v$ and the impact angle~$\phi$,
the solution is not unique and alternative fits of the SFD are possible.
Nevertheless, a reaccumulative event is expected at the origin of the Kalliope family.
This is closely related to the observed shape of (22)~Kalliope,
which is {\em aspherical\/} (Fig.~\ref{kalliope_meshlab}),
similarly as other M-type bodies.
Material rheology and non-zero friction during reaccumulation
must support all these topographic features.

%Basalt_4_60_2.2
%Basalt_6_45_2.6 w. Linus

According to \cite{Durda_2004Icar..170..243D},
Linus may be classified as a smashed-target satellite.
If we also include it
in the SFD, the collision should be even more energetic. 
Actually, Linus appears as an intermediate-sized fragment and 
its volume represents about twenty 10-km bodies.
It may thus be difficult to explain the existence of Linus
and the remaining fragments at the same time.

%%%%%%%%%%%%%%%%%%%%%%%%%%%%%%%%%%%%%%%%%%%%%%%%%%%%%%%%%%%%%%%%%%%%%%%%

\section{Collisional evolution}\label{collisional}

We simulated the long-term collisional evolution with the Boulder code
\citep{Morbidelli_2009Icar..204..558M}.
The collisional probabilities and impact velocities
for the relevant populations (main belt, Kalliope family)
were computed as follows:
\begin{center}
\begin{tabular}{lll}
MB--MB             & $2.86\cdot10^{-18}\,{\rm km}^{-2}\,{\rm y}^{-1}$ & $5.77\,{\rm km}\,{\rm s}^{-1}$ \\
MB--Kalliope       & $3.17$ & $5.58$ \\
Kalliope--Kalliope & $5.80$ & $4.75$ \\
\end{tabular}
\end{center}
The scaling law is similar to \cite{Benz_Asphaug_1999Icar..142....5B}
with lower strength at $D \simeq 100\,{\rm m}$
in order to match the observed SFD of the main belt, namely:
\begin{equation}
Q = Q_0 R^a + B\rho R^b\,,
\end{equation}
where
$Q_0 = 9\cdot 10^7\,{\rm erg}\,{\rm g}^{-1}$,
$a = -0.53$,
$B = 0.5\,{\rm erg}\,{\rm cm}^{-3}$,
$b = 1.36$, and
$R = D/2$ (in cm).
For simplicity, we also assumed the same density $\rho = 4.1\,{\rm g}\,{\rm cm}^{-3}$
as for (22) Kalliope (for ADAM shape model; \citealt{Ferrais_2022}),
but if it is differentiated, it may be more logical to assume a lower density
for fragments, corresponding to silicates ($\rho \simeq 3\,{\rm g}\,{\rm cm}^{-3}$).
On the other hand, for (22) itself, the value of~$Q$ might be larger than nominal,
if it is differentiated.

We accounted for a size-dependent dynamical decay,
as described in \cite{Cibulkova_2014Icar..241..358C}.
The decay in the pristine zone is relatively fast,
given the distance between the 5:2 and 7:3 resonances (0.14\,au).
Compared to the inner main belt (0.4\,au).
we expect it to be about 3 to 5 times faster.
For Linus, we artificially increased its lifetime,
as it is bound to (22) Kalliope.
Because the family itself must have been extended even at $t = 0$,
some bodies were initially close to or inside the resonances
(as in Sec.~\ref{orbital}).

Initial conditions for the main belt were close to the observed SFD,
because it is close to a steady state.
On contrary, we assumed a steep SFD for the synthetic family,
even below $D < 5\,{\rm km}$,
because there is no apparent reason, why it should be so shallow.
Moreover, there is other source of material we should not forget ---
28-km Linus, which is a regular intermediate-sized family member.
We included Linus in the SFD, because secondary collisions with Linus may contribute to the SFD.

Our results are plotted in Fig.~\ref{37_MB_Kalliope_LINUS_sfd_0900}.
Each simulation was run 10 times, in order to account for
stochasticity of collisions.
If the initial conditions correspond to a smooth power-law
with the slope $q = -3.0$,
it is impossible to explain the break at 5\,km
as well as non-existence of $D > 10$-km bodies.
To fit both features,
the SFD must be initially without $D > 10$-km bodies,
which actually creates the break at 5\,km
in the course of collisional evolution (Fig.~\ref{37_MB_Kalliope_LINUS_sfd_0900}, middle).
The minimum age of the family is then 800\,My.

Alternatively, if the initial SFD is scaled up by a factor of~5,
with an extremely steep slope $q = -10$ at the large-size end
and a shallow slope afterwards,
the whole SFD simply evolves downwards
and ends up similar as before;
age may reach up to 3.4\,Gy.
However, it is unlikely that the family is so old,
because its initial SFD is so extreme.
In 10\,\% of such simulations, Linus experienced a~catastrophic collision
which would be in contradiction with its very existence.

%%%%%%%%%%%%%%%%%%%%%%%%%%%%%%%%%%%%%%%%%%%%%%%%%%%%%%%%%%%%%%%%%%%%%%%%

\section{Orbital evolution}\label{orbital}

We simulated the long-term orbital evolution with the numerical integrator SWIFT
\citep{Levison_Duncan_1994Icar..108...18L},
supplemented with the Yarkovsky effect, YORP effect, collisional reorientations, or mass shedding
\citep{Broz_2011MNRAS.414.2716B}.

\begin{figure}
\centering
\includegraphics[width=8.5cm]{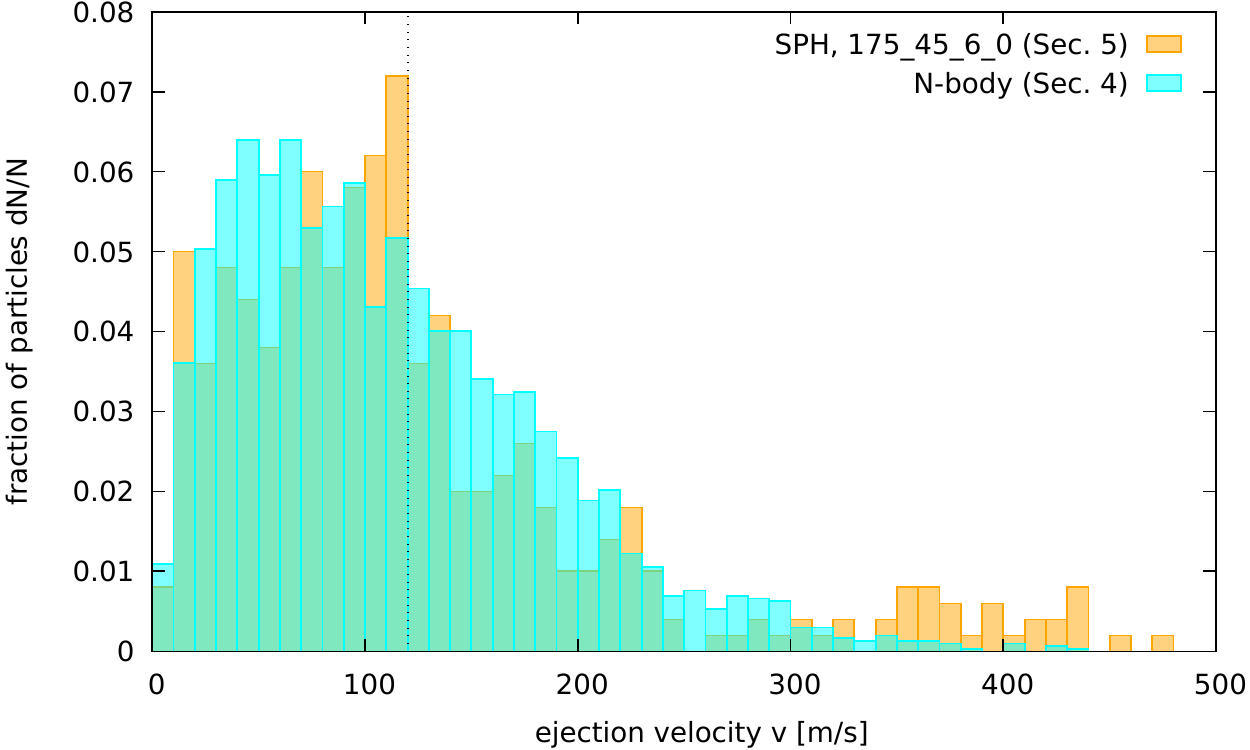}
\caption{Histogram of the ejection speed in our orbital model (blue), compared with one SPH simulation from Sec.~\ref{sph},
{\tt 175\char`_45\char`_6\char`_0} (orange).
The escape speed is indicated by the dotted line.}
\label{kalliope13_hist_velocity}
\end{figure}

Our synthetic Kalliope family was created as an artificial
breakup with an assumed velocity field.
It contains 10 times more bodies than the observed SFD,
to have enough km-sized bodies at late stages.
Velocity distribution is size-dependent, with 
$v_x(D) = 24\,{\rm m}\,{\rm s}^{-1}\,(D/5\,{\rm km})^{-1}$,
and similarly for other components.
The histogram of $|\vec v|$ exhibits a characteristic peak at the escape velocity
and it is similar to the outcomes of SPH simulations
(see Fig.~\ref{kalliope13_hist_velocity} and Sec.~\ref{sph}).
The field is isotropic in the Cartesian space.
Of course, in the osculating element space,
the distribution is no longer isotropic ---
it is rather given by the impact geometry, namely
the true anomaly
$f = 100^\circ$
and the argument of pericentre
$\omega = 330^\circ$,
so that iso-velocity ellipses resemble the core of the family
(as in Fig.~\ref{aei2_vgauss}).

\begin{figure*}
\centering
\begin{tabular}{@{}c@{}c@{}c@{}}
initial conditions & 200\,My & 1\,Gy \\
\includegraphics[width=6.1cm]{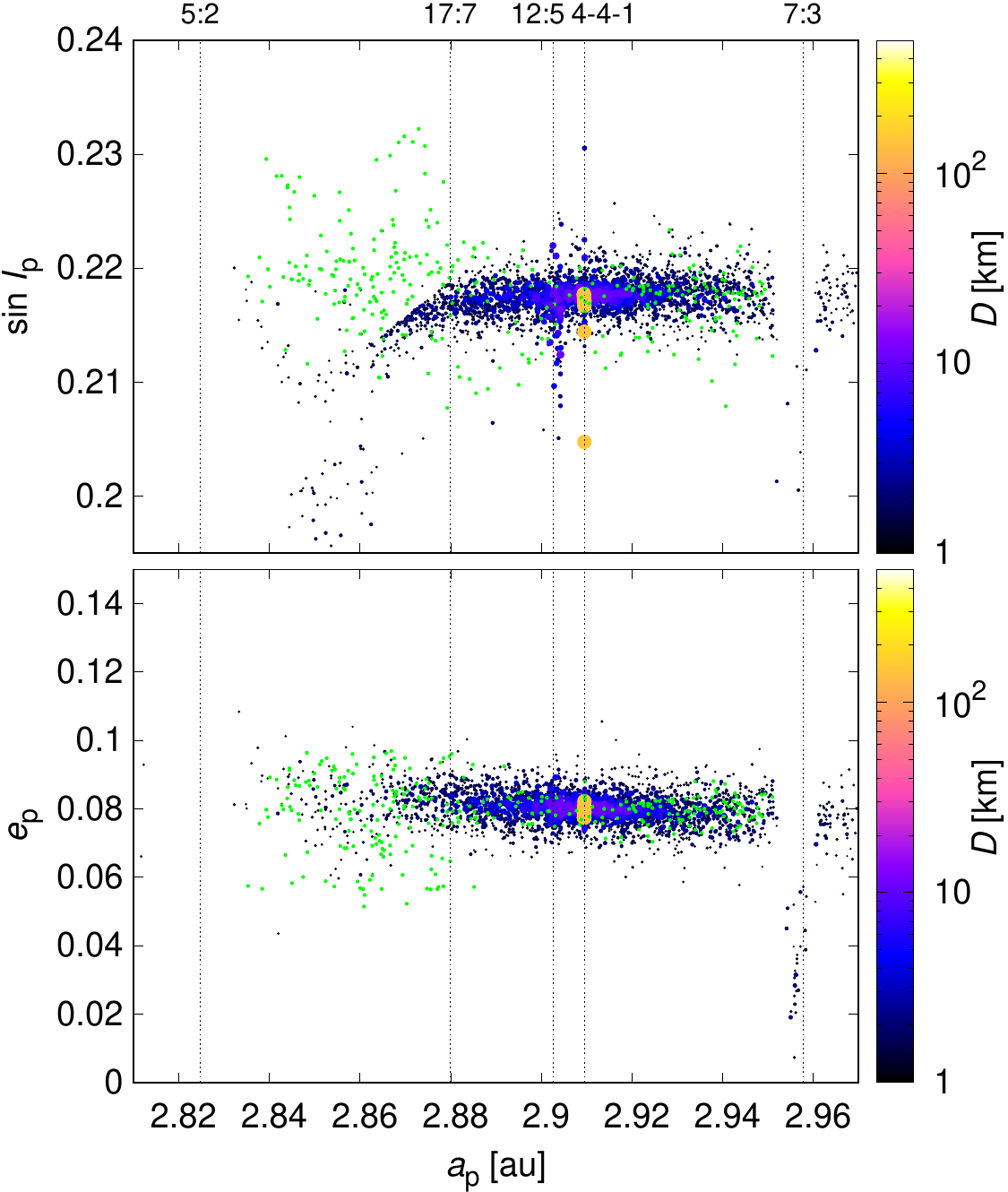} &
\includegraphics[width=6.1cm]{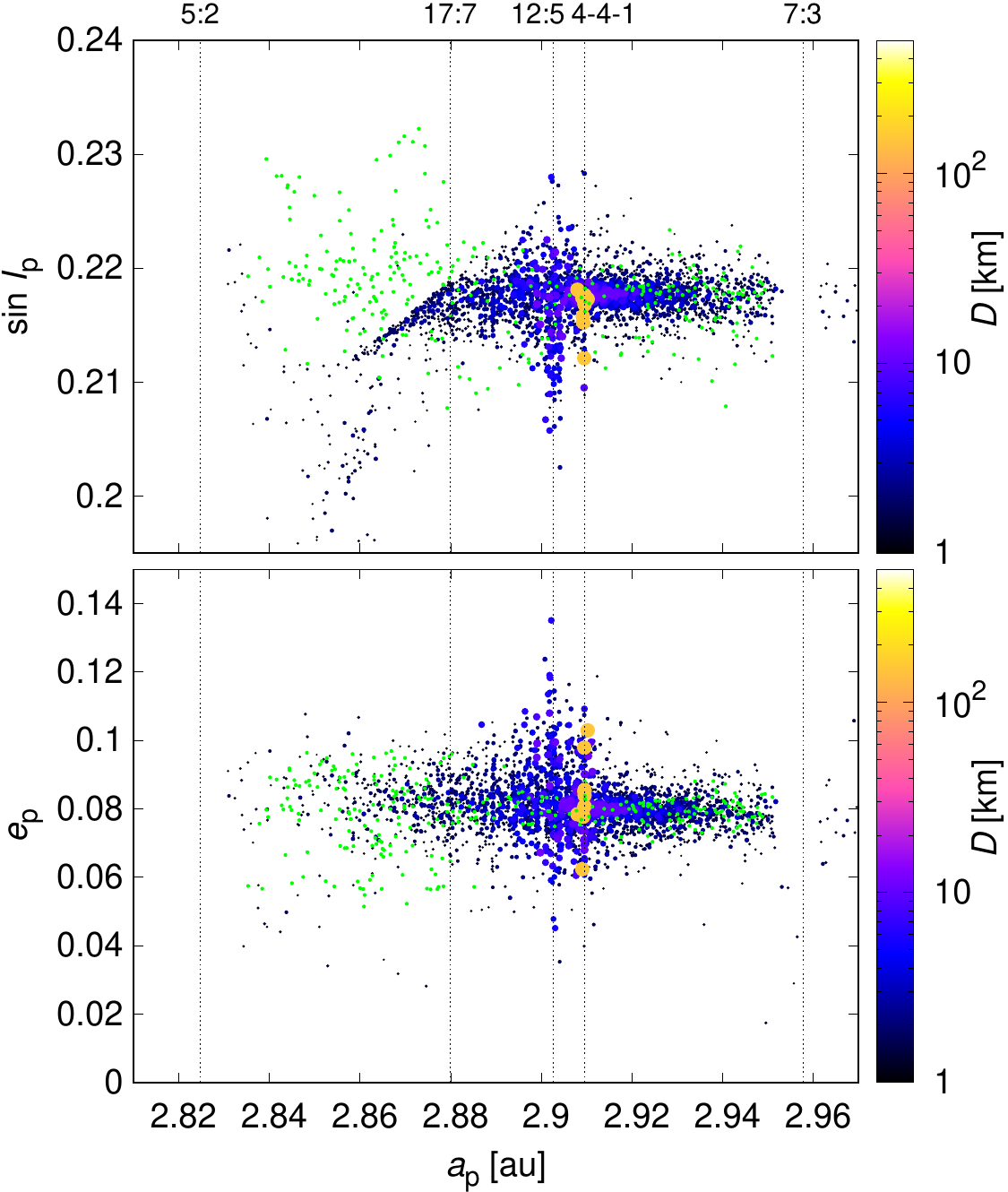} &
\includegraphics[width=6.1cm]{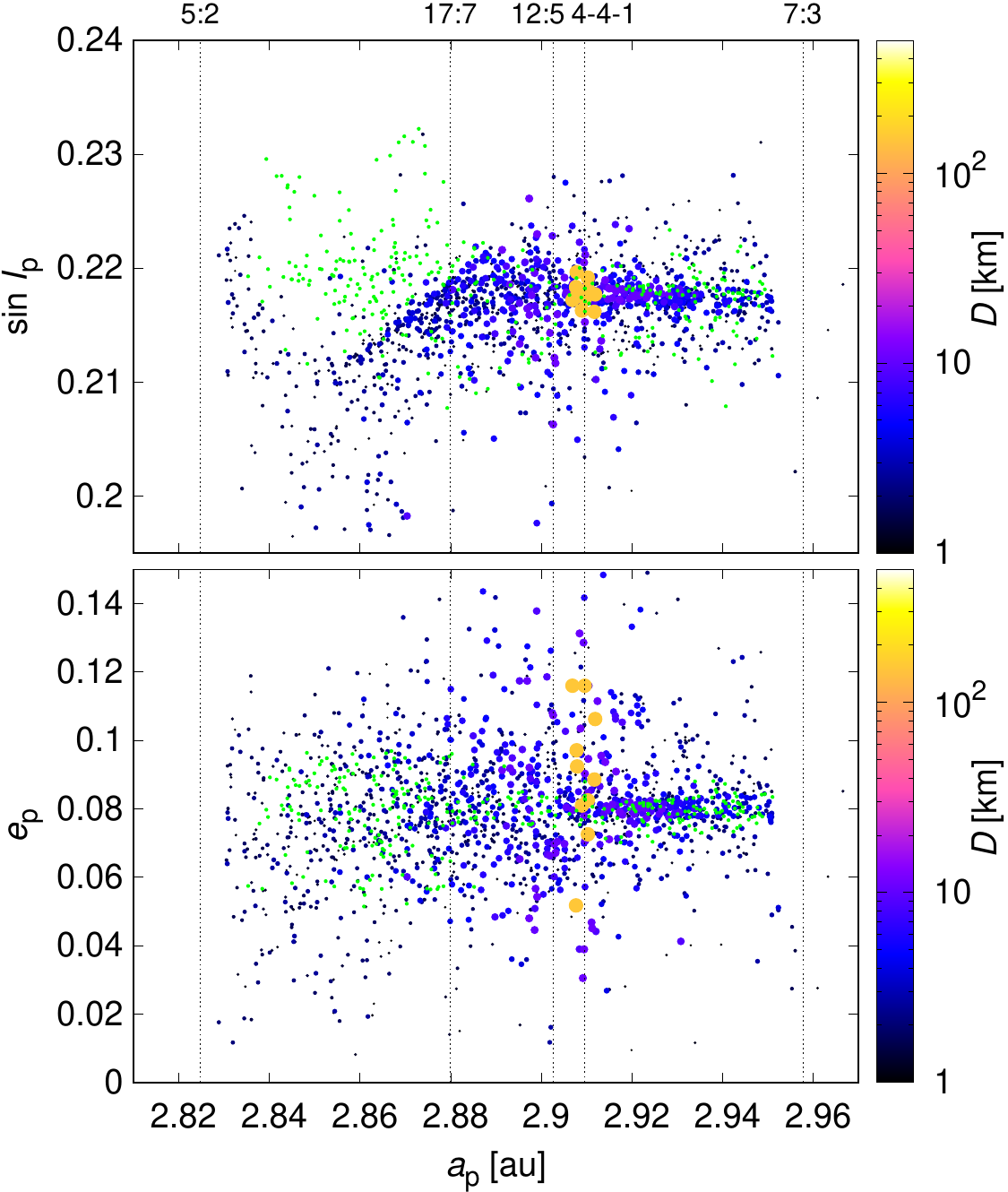} \\
\end{tabular}
\caption{
Orbital evolution of the Kalliope family.
Proper semimajor axis $a_{\rm p}$ vs. proper eccentricity $e_{\rm p}$ (bottom)
as well as proper inclination $\sin I_{\rm p}$ are plotted
in the course of time:
initial conditions (left), 200\,My (middle), and 1\,Gy (right).
Colours and symbols correspond to the actual diameters.
Observed family is plotted for comparison (green).
The number of synthetic bodies is 10 times larger (3020 vs. 302) to improve the statistics.
Evolution is driven by the Yarkovsky effect and mean-motion resonances,
especially 12:5 and $4{-}1{-}1$;
(22)~Kalliope (yellow) is indeed affected by the latter.
}
\label{kalliope-1}
\end{figure*}

The time step was $\Delta t = 36.525\,{\rm d}$,
the time span $t_2 - t_1 = 1\,{\rm Gy}$.
Mean elements were computed using convolution filters
\citep{Quinn_1991AJ....101.2287Q},
with the input sampling 1\,y,
filters A, A, A, B
decimation factors 10, 10, 10, 3,
and the output sampling 3000\,y.
Proper elements were determined by the frequency-modified Fourier transform
\citep{Sidlichovsky_1996CeMDA..65..137S}
from 1024 samples,
and the output sampling was 0.1\,My.
This output is mostly compatible with other types of proper elements;
a minor difference was apparent for proper inclinations,
below $a_{\rm p} = 2.88\,{\rm au}$
An artefact ($\sin I_{\rm p}$ lower by 0.02 in Fig.~\ref{kalliope-1})
may occur if $g$ or $s$ frequencies of orbits approaching the 5:2
resonance become too close to the passbands of our digital filters.
A solution would be to use a modified setup (e.g., A, A, B, B),
nevertheless, other parts of the proper elements space were not affected.

Our results are plotted in Fig.~\ref{kalliope-1}.
Initially, the synthetic family extended across the pristine zone
in semimajor axis, due to outliers in the velocity field.
On the contrary, is was narrow in eccentricity and inclination,
similarly to the core of the observed family.
(22) Kalliope and all its clones were located in the $4{-}1{-}1$ resonance.

In the course of evolution, the number of bodies decreases
due to the Yarkovsky drift, perturbations by the 5:2 and 7:3 resonances
and scattering by Jupiter;
the exponential decay time scale $\tau$ is approximately $1.2\,{\rm Gy}$.
The synthetic family becomes more spread not only in $a$, but also in $e$, $I$,
due to the 17:7, 12:5 and $4{-}1{-}1$ resonances.
Interestingly, (22) Kalliope and its clones chaotically diffuse,
which offers an opportunity to determine the age independently.
For the time $t \lesssim 500\,{\rm My}$, the number of bodies below
the 17:7 resonance and their spread is insufficient.
For the time $t \gtrsim 1\,{\rm Gy}$, majority of (22) clones is
spread by more than 0.01 in $e$.
Taken together, the age of the family must be $(750\pm 250)\,{\rm My}$.
This is fully compatible with the collisional evolution
(Fig.~\ref{37_MB_Kalliope_LINUS_sfd_0900}, middle).
Actually, it also confirms that a hypothesis of a 3.4\,Gy-old family 
(Fig.~\ref{37_MB_Kalliope_LINUS_sfd_0900}, right) is excluded.
More precise age determination is not possible due to the limited
number of observed bodies and systematic uncertainty of the density
of ejected fragments.
This is closely related to the internal structure of the parent body.

%%%%%%%%%%%%%%%%%%%%%%%%%%%%%%%%%%%%%%%%%%%%%%%%%%%%%%%%%%%%%%%%%%%%%%%%

\section{SPH simulations}\label{sph}

We simulated a break-up of the Kalliope parent body by means of
the smoothed-particle hydrodynamics (SPH), with the Opensph solver
\citep{Sevecek_2019A&A...629A.122S,Sevecek_2019ascl.soft11003S}.
A principal question is: `Is differentiated different from homogeneous?'
Of course, the body has a certain density profile $\rho(r)$ (cf.~\citealt{Ferrais_2022}),
but hereinafter we are interested in $\rho$ of the ejecta,
or in the chemical composition (silicates vs. metal).
According to an analogy with the Earth-Moon system,
we would expect that the moon should have lower~$\rho$,
corresponding to the mantle, not to the core \citep{Canup_2014RSPTA.37230175C}.
The same is true for other ejecta.

Another principal question is:
how much ejecta must be ejected (to $\infty$) in order to form a massive moon
on a bound orbit (to ${\ll}\,\infty$)?
In other words, is the observed Kalliope family compatible with Linus?

Initial conditions of our simulation are somewhat simplified.
We assumed the target is either spherical or Maclaurin ellipsoid,
i.e., either static or rotating.
The interior was differentiated,
composed of a metallic core ($8\,{\rm g}\,{\rm cm}^{-3}$)
and a silicate mantle ($3\,{\rm g}\,{\rm cm}^{-3}$),
or alternatively homogeneous (for comparison).
Depending on the structure,
we expected substantial differences in
shock wave propagation,
reflection,
attenuation,
focussing,
ejection of material,
etc.
The spherical target diameter $D = 153\,{\rm km}$,
the projectile diameter~$d$ was varied.
The core--mantle boundary was adjusted so that 
the total mass and volume correspond to the observed values \citep{Ferrais_2022};
the core diameter was then $D_{\rm c} = 92\,{\rm km}$.
We expected that low- and mid-energy collisions will not eject too much mass.
High-energy collisions may require a somewhat larger target.
Moreover, they may potentially explain a high density of the remnant,
if enough mantle material is ejected.
The initial specific internal energy was low ($10^3\,{\rm J}\,{\rm kg}^{-1}$),
constant throughout the interior.
The targets were put in hydrostatic equilibrium before impact.

Similar computations were made for Maclaurin ellipsoids.
We assumed the current rotation period,
$P \doteq 4.1482\,{\rm h}$,
and estimated the eccentricity from ($\omega = 2\pi/P$):
\begin{equation}
{\omega\over\pi G\rho} = 2{\sqrt{1-e^2}\over e^3} (3-2e^2)\arcsin e - {6\over e^2} (1-e^2)\,.
\end{equation}
Therefore, the core is less eccentric than the mantle.

We also varied the impact velocity $v_{\rm imp}$,
between $5$ and $ 6\,{\rm km}\,{\rm s}^{-1}$,
and the impact angle $\phi_{\rm imp}$,
from $0$ to $45^\circ$.
In the case of ellipsoids, the impacts were in the equatorial plane.
Materials were described by the
\cite{Tillotson_1962geat.rept.3216T} equation of state,
Drucker--Prager rheology,
Grady--Kipp fragmentation,
Weibull flaw distribution,
where principal parameters were taken from
\cite{Benz_Asphaug_1999Icar..142....5B,Maurel_2020Icar..33813505M}
and listed in Tab.~\ref{tab1}.
The yield strength was dependent on the internal energy.
The core material is modelled with the same strength model as the mantle
(but with different constants).

\begin{table}
\caption{Material parameters of the SPH simulations.}
\label{tab1}
\centering
\begin{tabular}{llll}
$\rho$         & $ 3              $ & $ 8             $ & ${\rm g}\,{\rm cm}^{-3}$ \\
$A$            & $ 26.7           $ & $ 128           $ & ${\rm GPa}$ \\  % Maurel (2020)
$B$            & $ 26.7           $ & $ 105           $ & ${\rm GPa}$ \\  % Maurel (2020)
$\mu$          & $ 22.7           $ & $  82           $ & ${\rm GPa}$ \\  % wiki
$E$            & $  8             $ & $ 211           $ & ${\rm GPa}$ \\  % wiki
$Y$            & $  3.5           $ & $   0.35        $ & ${\rm GPa}$ \\  % wiki
$C$            & $  0.09          $ & $   0.09        $ & ${\rm GPa}$ \\  % Brandes (1966)
$a$            & $ 0.5            $ & $ 0.5           $ & \\
$b$            & $ 1.5            $ & $ 1.5           $ & \\
$\alpha$       & $ 5              $ & $ 5             $ & \\
$\beta$        & $ 5              $ & $ 5             $ & \\
$U_{\rm melt}$ & $   3.4          $ & $ 1.0           $ & ${\rm MJ}\,{\rm kg}^{-1}$ \\  % Sugiura (2020; PhD thesis), matconst.dat
$U_{\rm iv}$   & $   4.72         $ & $ 1.42          $ & ${\rm MJ}\,{\rm kg}^{-1}$ \\  % matconst.dat
$U_{\rm cv}$   & $  18.2          $ & $ 8.45          $ & ${\rm MJ}\,{\rm kg}^{-1}$ \\  % matconst.dat
$U_{\rm subl}$ & $ 487            $ & $ 9.5           $ & ${\rm MJ}\,{\rm kg}^{-1}$ \\  % Tillotson (1962)
$\mu_{\rm i}$  & $ 2              $ & $ 2             $ & \\  % ???
$\mu_{\rm d}$  & $ 0.8            $ & $ 0.8           $ & \\  % ???
$k$            & $ 4\cdot 10^{35} $ & $ 1\cdot10^{23} $ & ${\rm m}^{-3}$ \\  % Maurel (2020)
$m$            & $ 9              $ & $ 8             $ & \\  % Maurel (2020)
\end{tabular}
\tablefoot{
$\rho$ denotes the density,
$A$~bulk modulus,
$B$~non-linear modulus,
$\mu$~shear modulus,
$E$~elastic modulus, 
$Y$~yield strength (for comparison),
$C$~cohesion,
$a$, $b$, $\alpha$, $\beta$ Tillotson parameters,
$U_{\rm melt}$~internal energy of melting (for comparison),
$U_{\rm iv}$~incipient vaporisation,
$U_{\rm cv}$~complete vaporisation,
$U_{\rm subl}$~sublimation,
$\mu_{\rm i}$~internal friction,
$\mu_{\rm d}$~dry friction,
$k$~Weibull coefficient,
$m$~Weibull exponent.
}
\end{table}

% https://en.wikipedia.org/wiki/Elastic_properties_of_the_elements_(data_page)
% https://en.wikipedia.org/wiki/Shear_modulus
%
% A  = 170    GPa ... bulk, volume
% mu =  82    GPa ... shear, rigidity
% E  = 211    GPa ... elastic, Young, tensile
% T  = 540    GPa ... tensile strength, maximum
% Y  =   0.35 GPa ... yield strength, plasticity
% C  =   0.09 GPa ... cohesion 

% T_melt = 1811 K
% c_v = c_p = 450 J kg^-1 K^-1
% u = c_v*T = 0.8 MJ kg^-1
%
% T_boil = 3134 K
% c_p > c_v
% c_p = c_v + R
% R   =   8 J kg^-1 K^-1

\begin{figure*}
\centering
\begin{tabular}{@{}c@{}c@{}c@{}c@{}}
%\verb|kalliope4| &
%\verb|kalliope3| &
%\verb|kalliope14| &
%\verb|kalliope9| \\
\verb|153_29_6_45| &
\verb|153_29_6_45| &
\verb|165_45_5_30| &
\verb|175_45_5_30| \\
sphere homogeneous &
sphere differentiated &
Maclaurin homogeneous &
Maclaurin differentiated \\[.1cm]
\includegraphics[width=4.6cm]{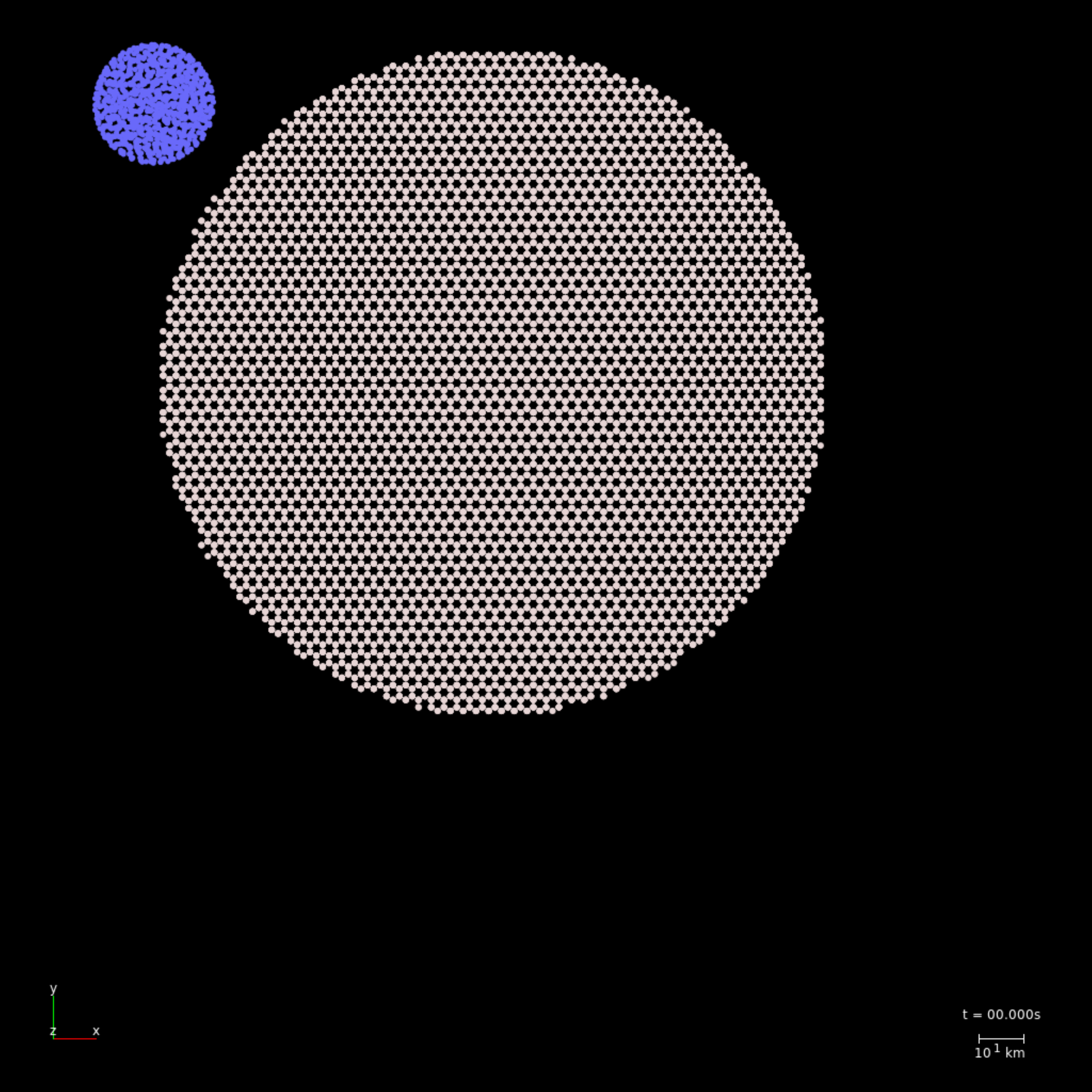} &
\includegraphics[width=4.6cm]{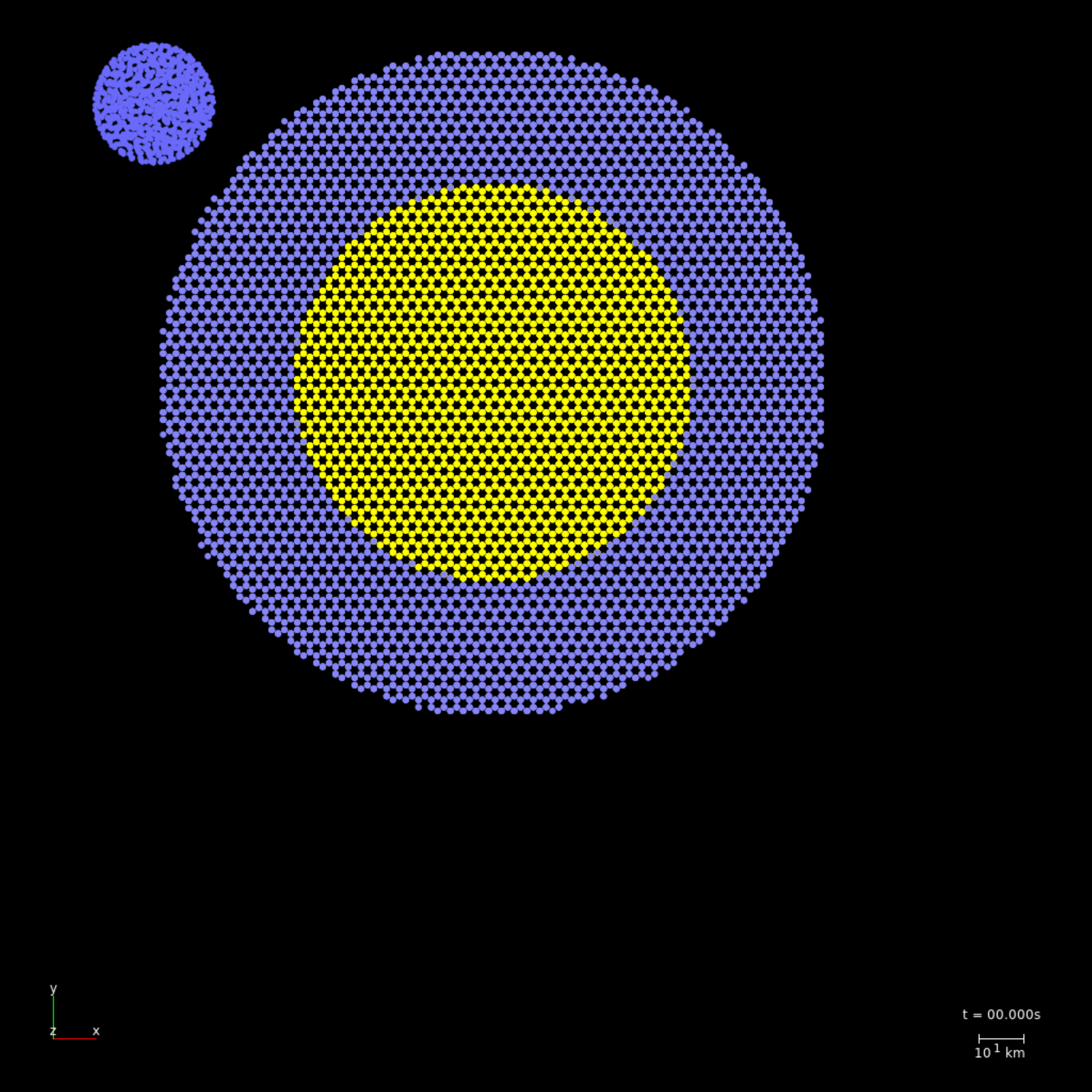} &
\includegraphics[width=4.6cm]{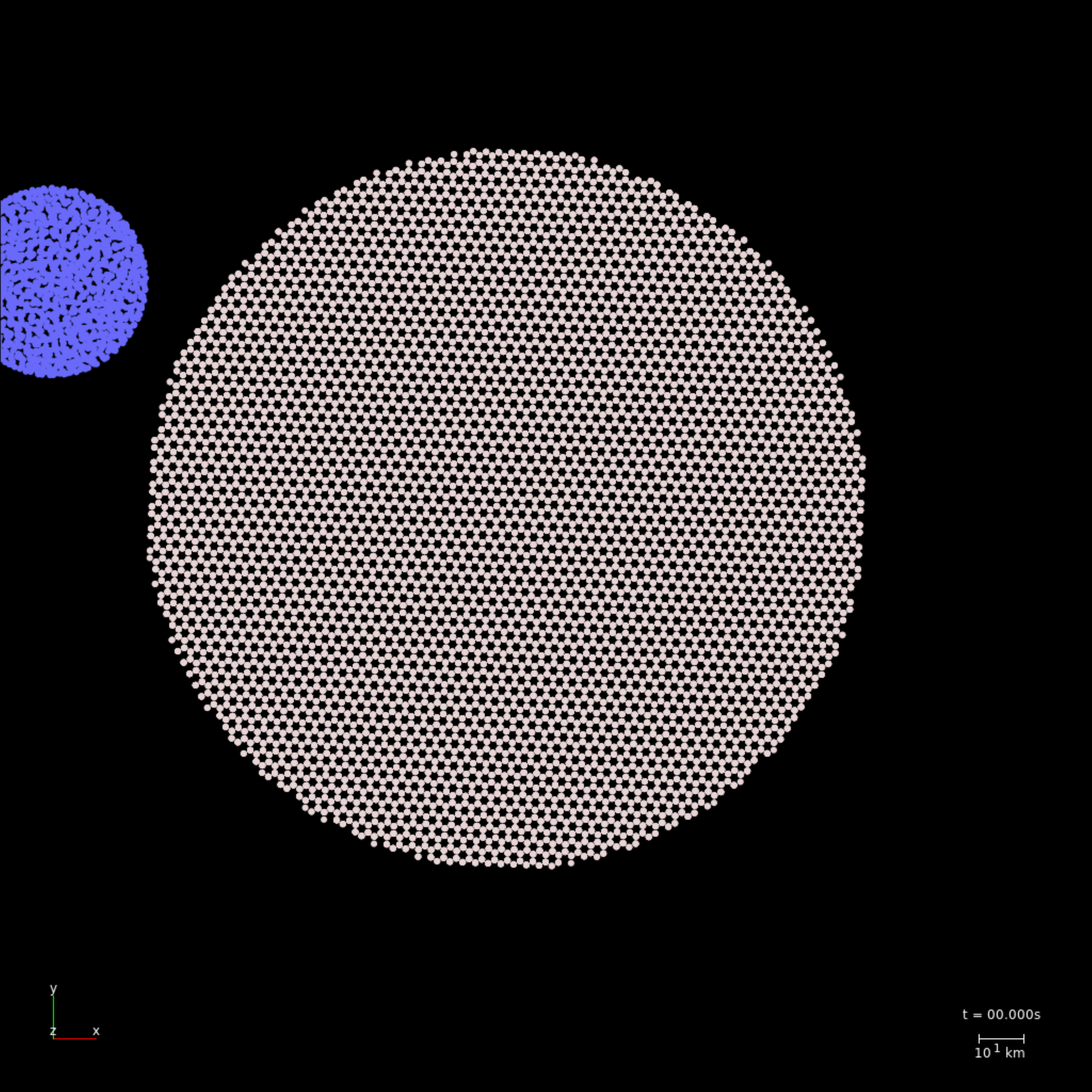} &
\includegraphics[width=4.6cm]{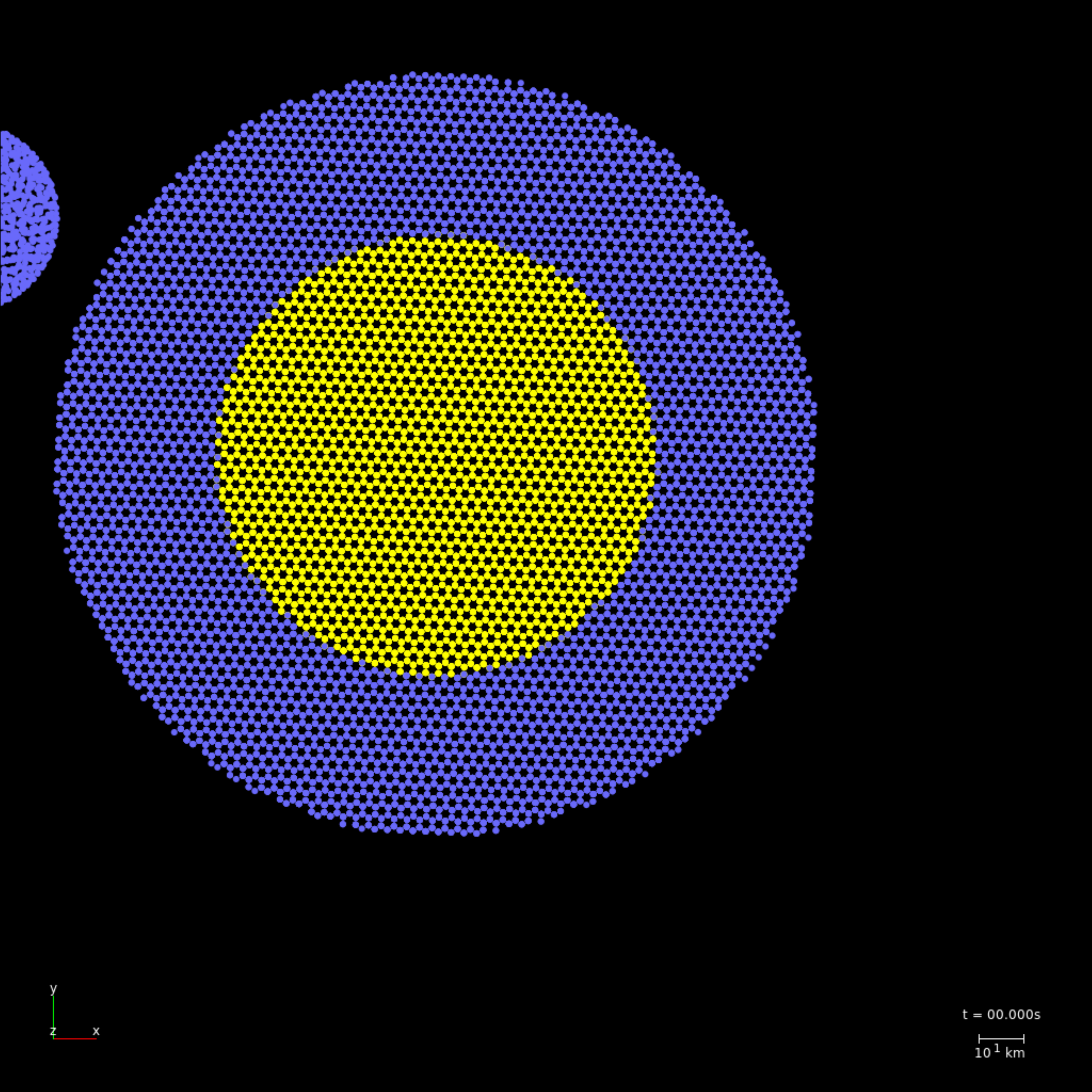} \\[.1cm]
\includegraphics[width=4.6cm]{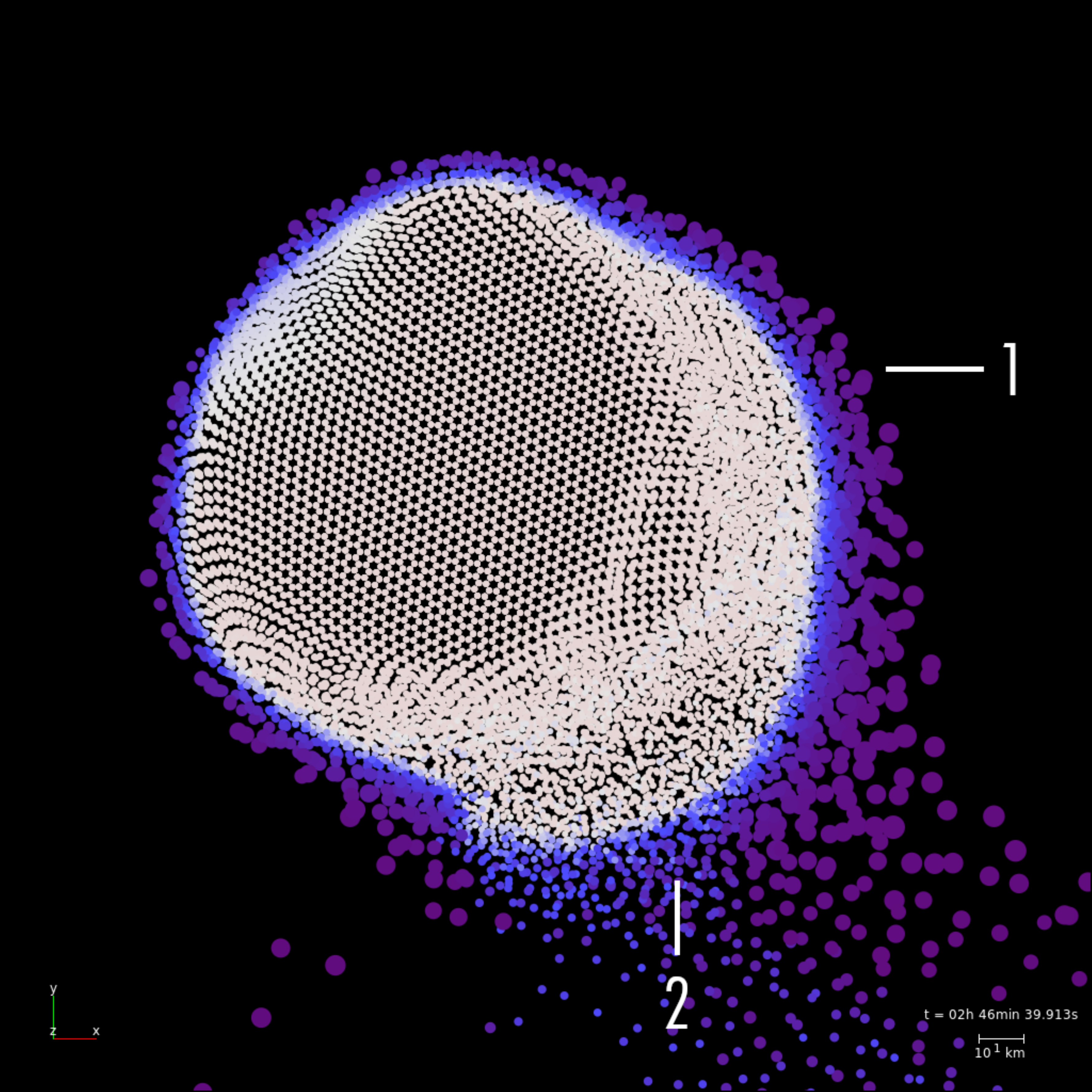} &
\includegraphics[width=4.6cm]{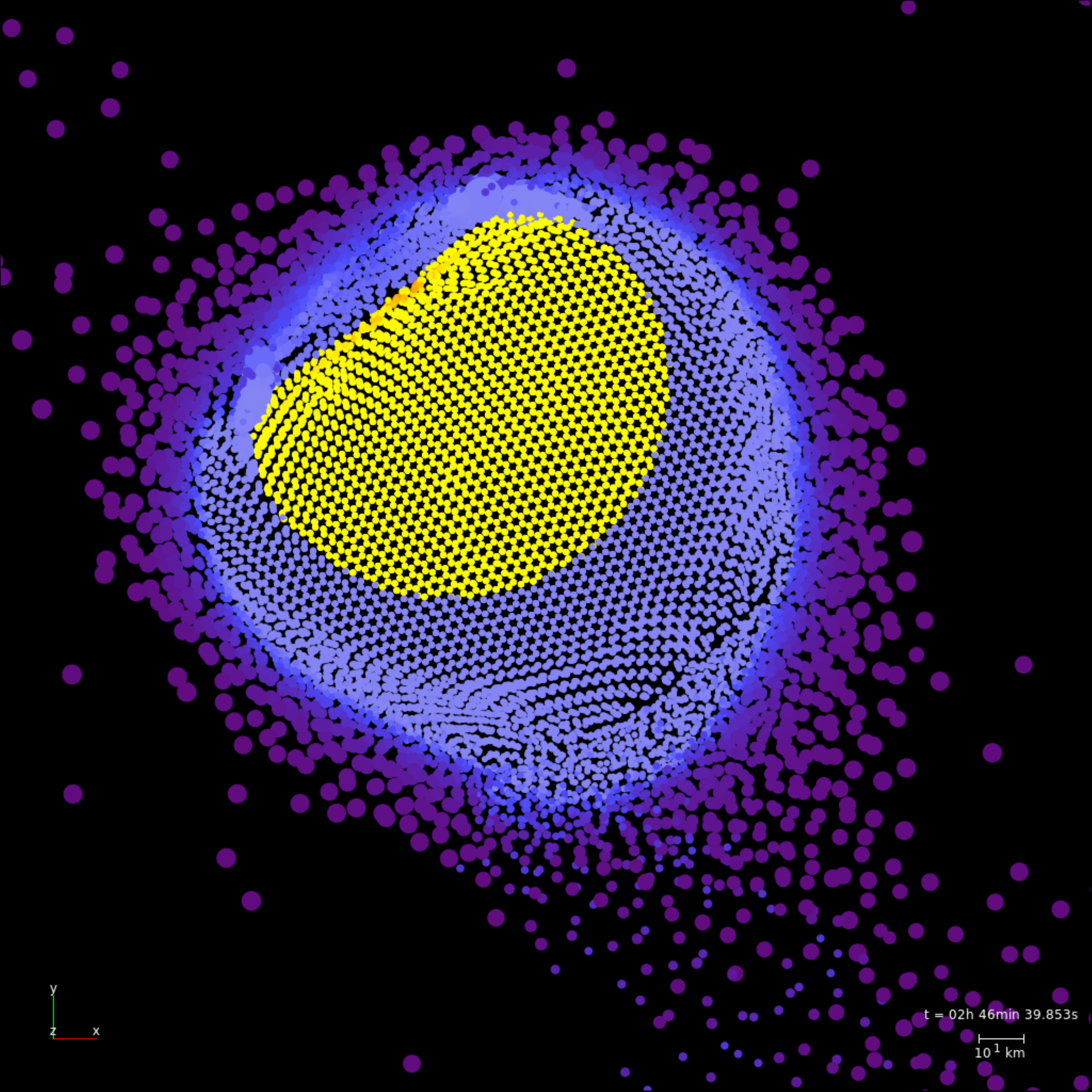} &
\includegraphics[width=4.6cm]{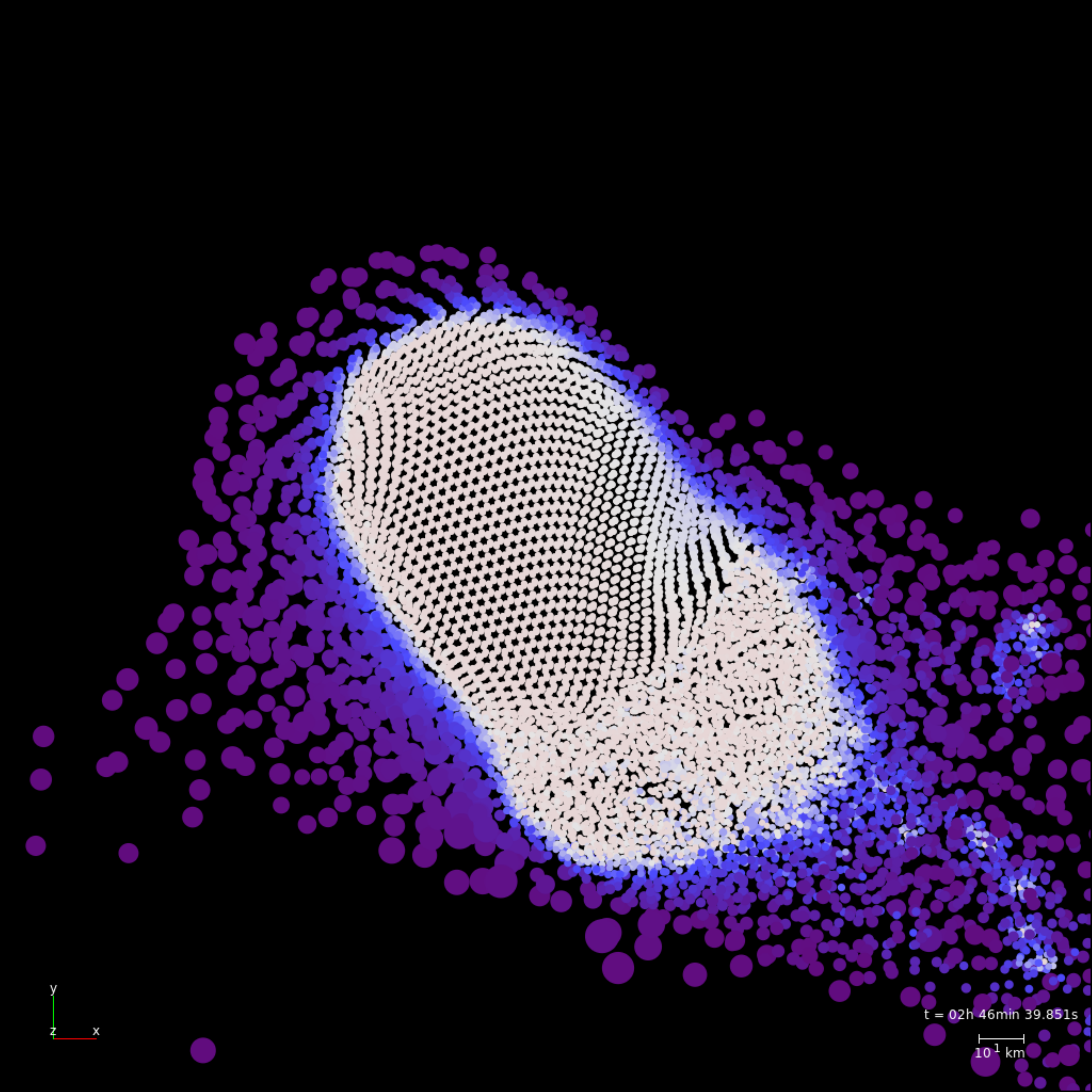} &
\includegraphics[width=4.6cm]{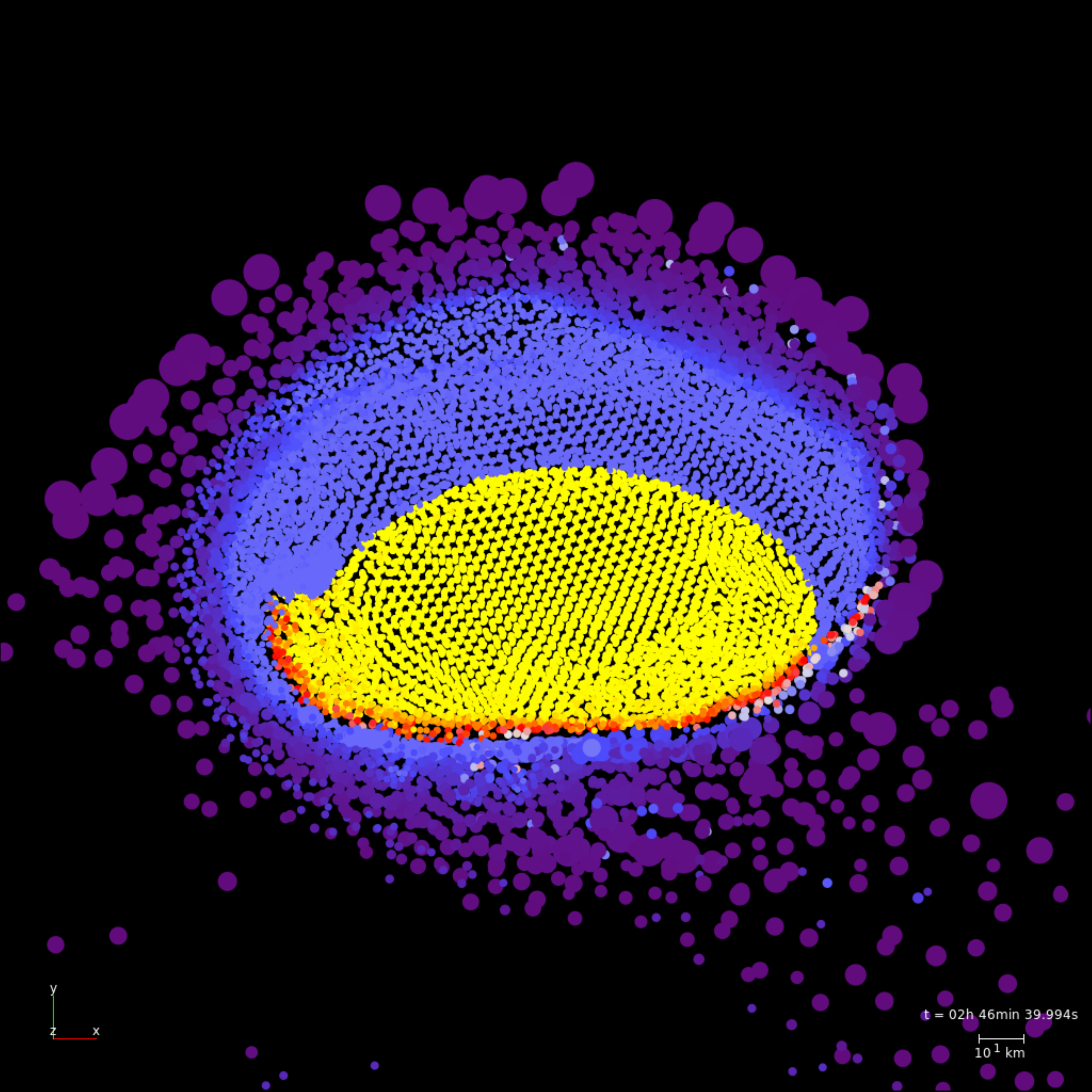} \\[.1cm]
\includegraphics[width=4.6cm]{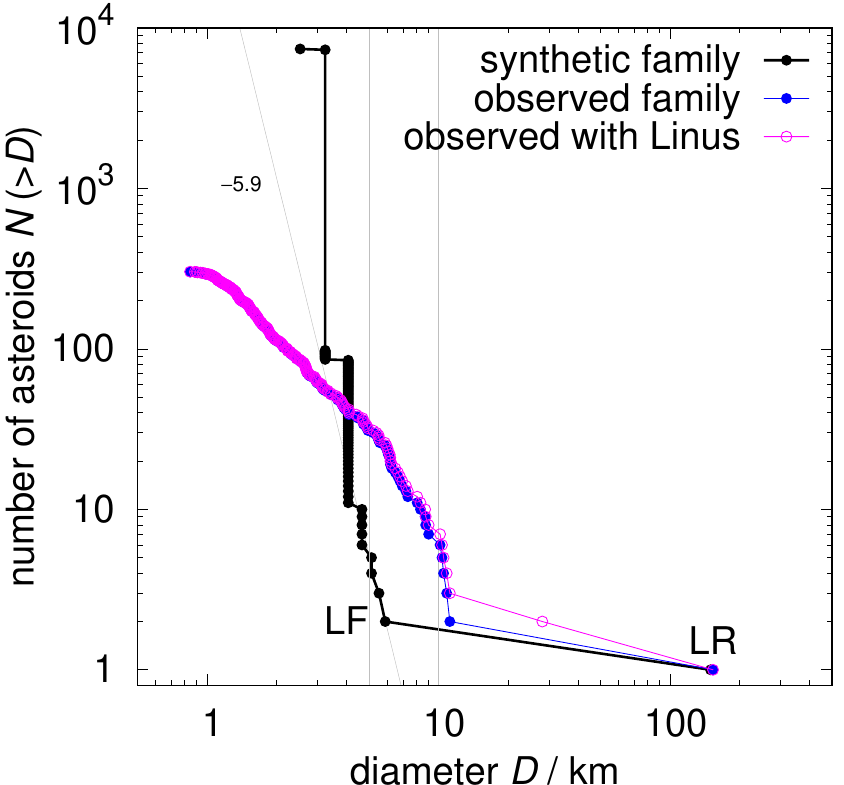} &
\includegraphics[width=4.6cm]{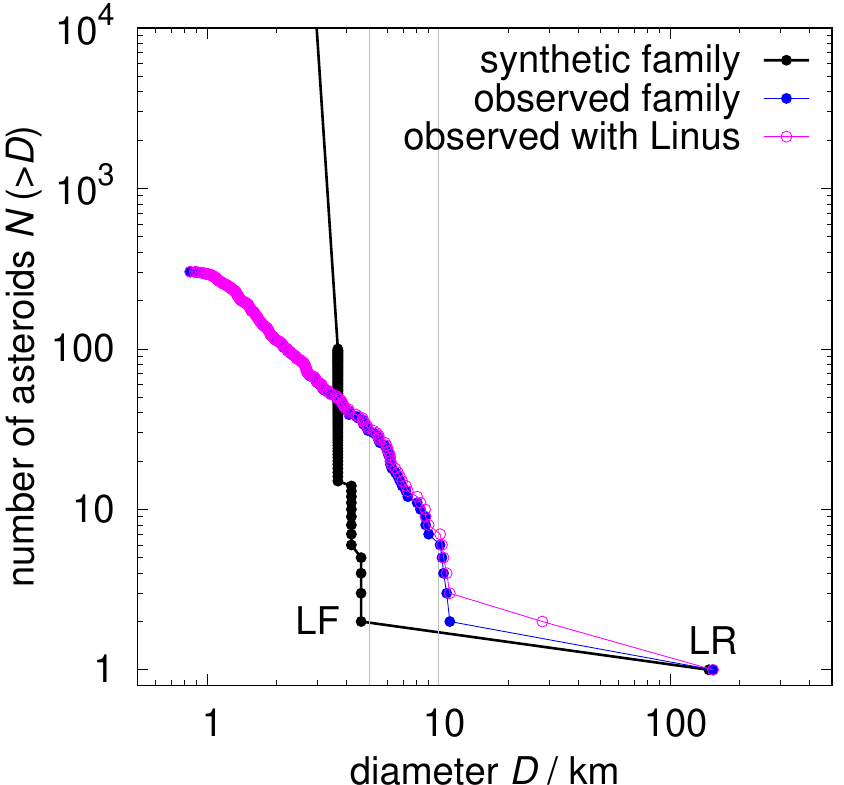} &
\includegraphics[width=4.6cm]{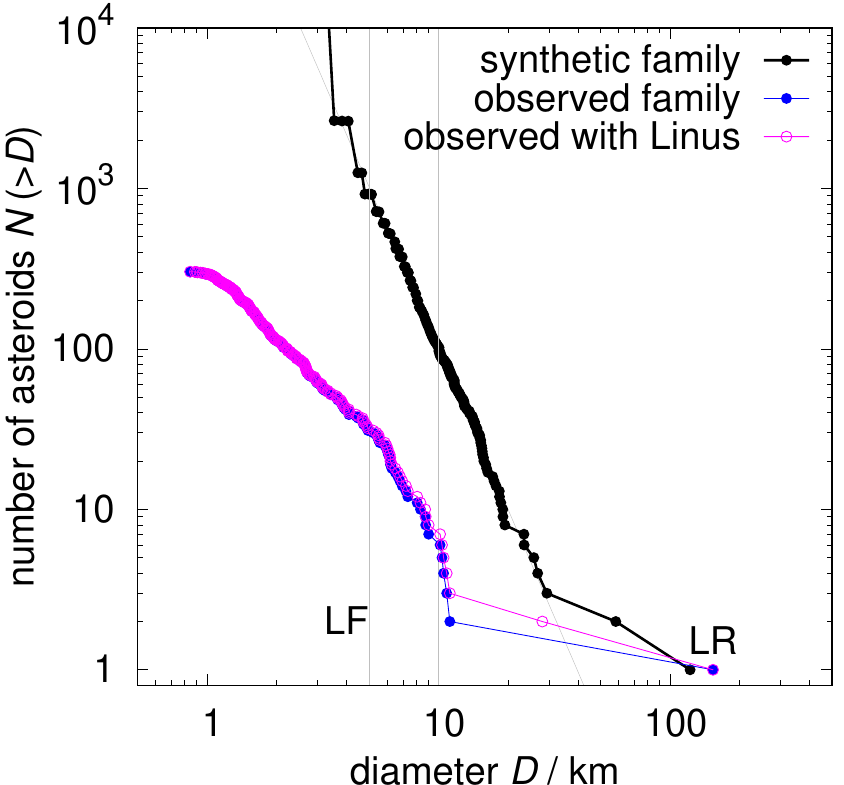} &
\includegraphics[width=4.6cm]{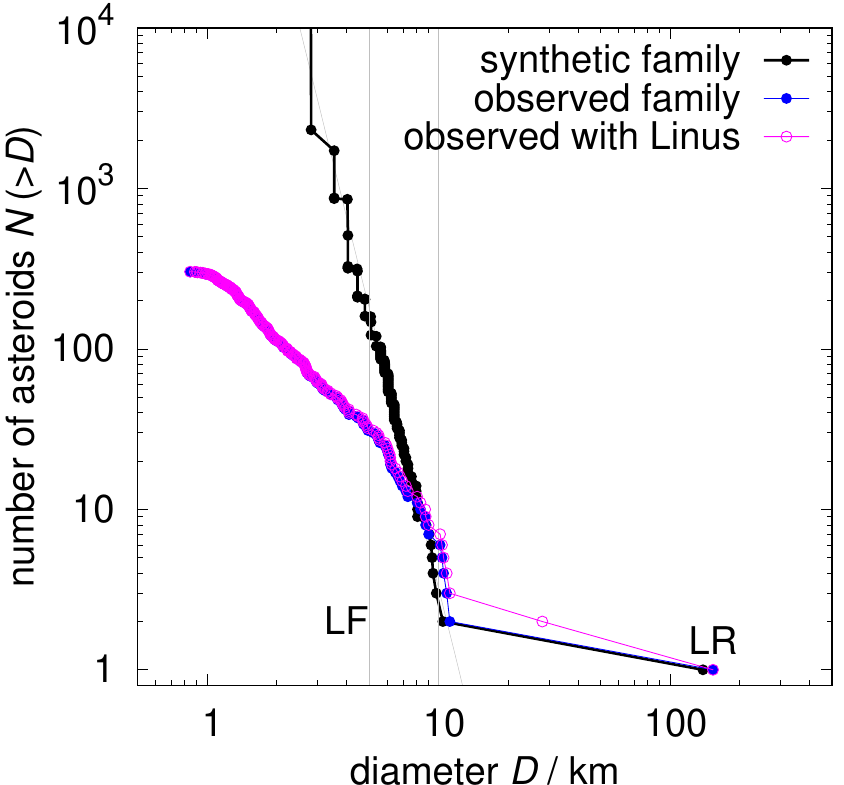} \\[.1cm]
UNDERSHOOT &
UNDERSHOOT &
OVERSHOOT &
NO LINUS \\
\end{tabular}
\caption{SPH simulations and resulting size-frequency distributions of the Kalliope family:
initial conditions (0\,s; top),
end of the fragmentation phase (10\,000\,s; middle),
end of the reaccumulation phase (300\,000\,s; bottom).
We tested different initial conditions:
homogeneous sphere (1st column),
differentiated sphere (2nd),
homogeneous Maclaurin ellipsoid (3rd),
differentiated Maclaurin ellipsoid (4th).
The title ({\tt XXX\char`_YY\char`_Z\char`_ŽŽ}) corresponds to
the target diameter~(in~km),
projectile diameter~(km),
impact speed~(km/s), and
impact angle~(deg).
We only plot a subset of SPH particles having $|z| < 10\,{\rm km}$
to have a clear view of the interior.
Their colours correspond to the density~$\rho$, on the scale:
violet (${\sim}\,0$),
blue (2.7),
white (4.1),
yellow ($8\,{\rm g}\,{\rm cm}^{-3}$).
Animations are available at
\url{https://sirrah.troja.mff.cuni.cz/~mira/kalliope/}.
}
\label{kalliope4_homo_DP_153_29_6_45_size_distribution}
\end{figure*}

\setcounter{figure}{5}
\begin{figure*}
\centering
\begin{tabular}{@{}c@{}c@{}}
%\verb|kalliope10| &
%\verb|kalliope13| \\
\verb|165_45_6_0| &
\verb|175_45_6_0| \\
Maclaurin homogeneous &
Maclaurin differentiated \\[.1cm]
\includegraphics[width=4.6cm]{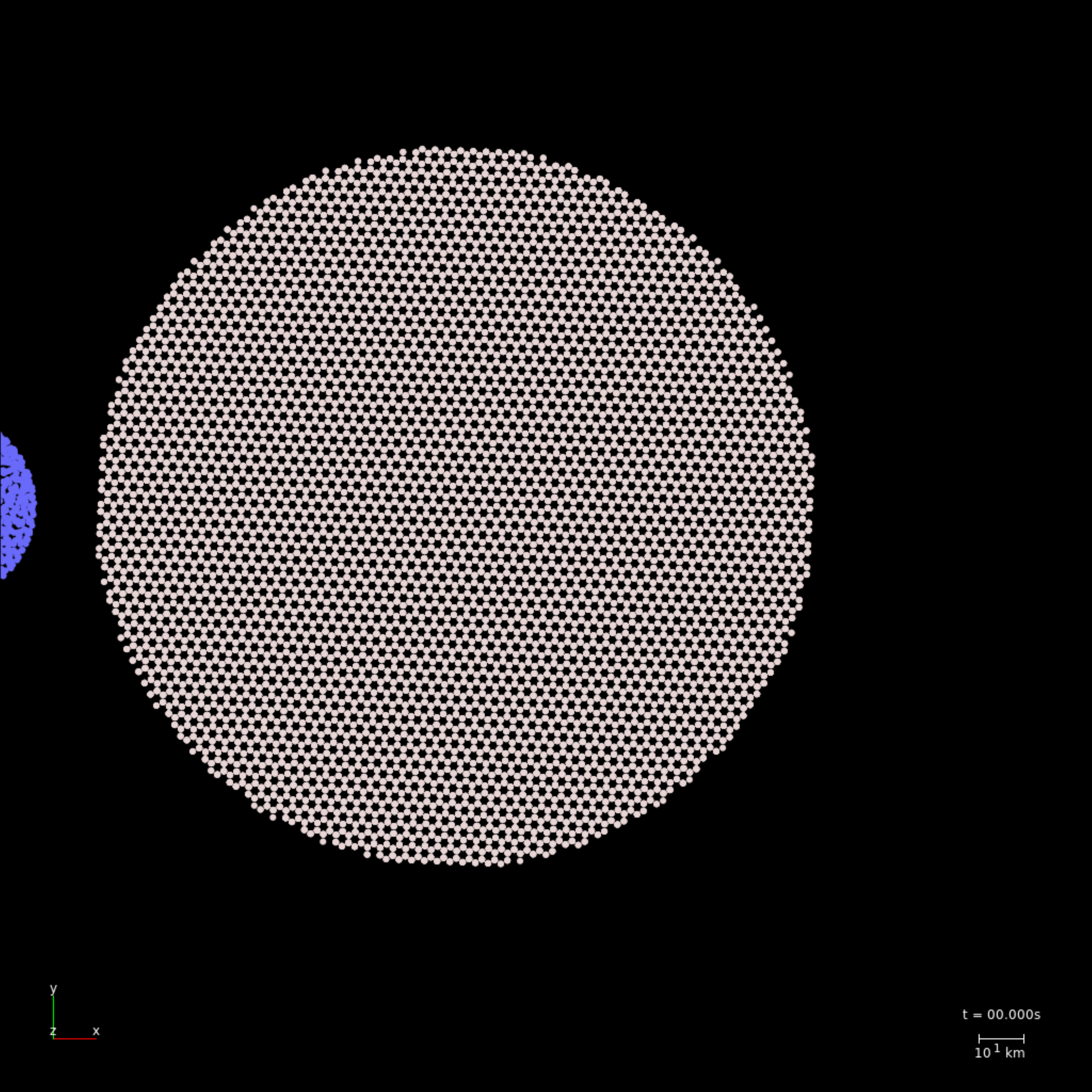} &
\includegraphics[width=4.6cm]{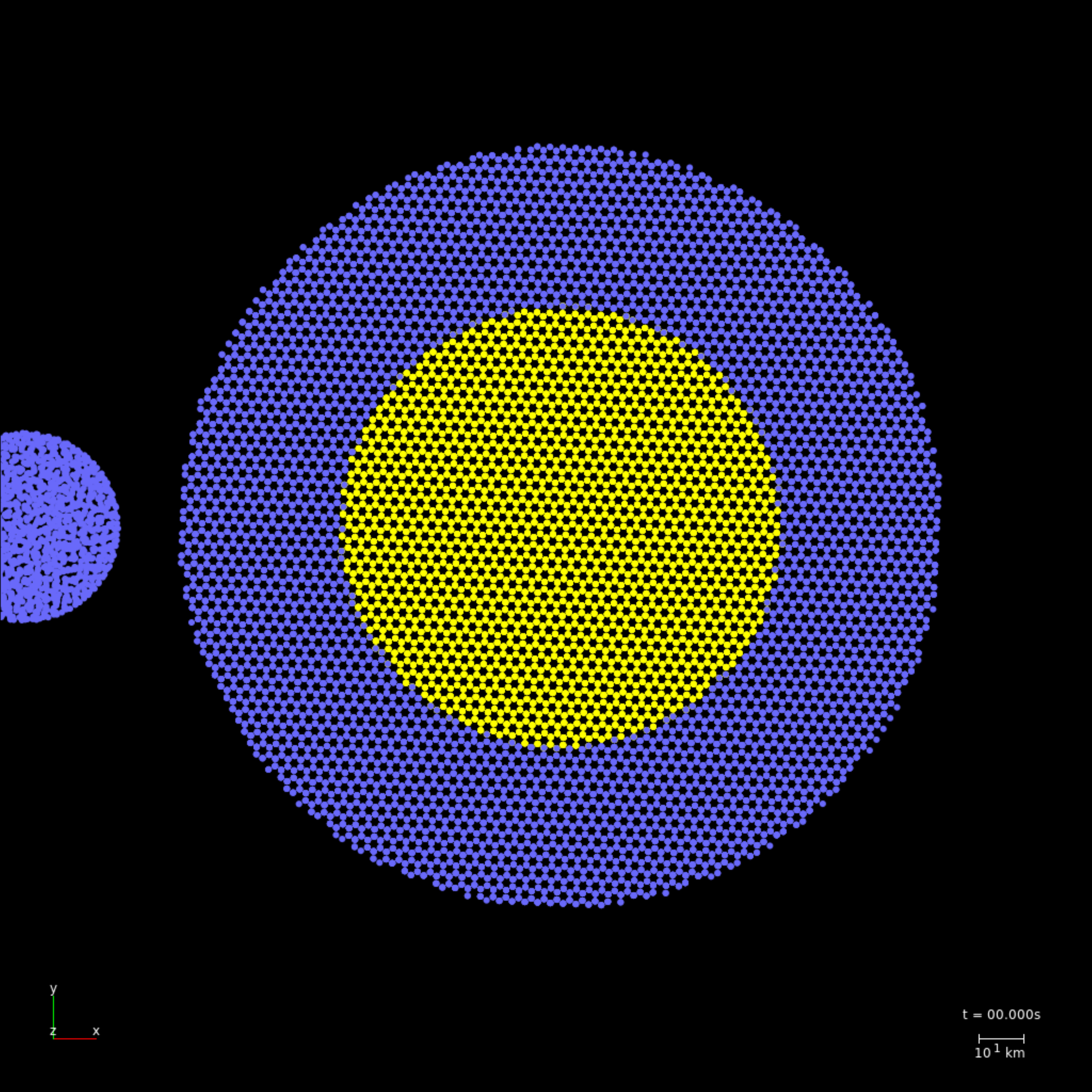} \\[.1cm]
\includegraphics[width=4.6cm]{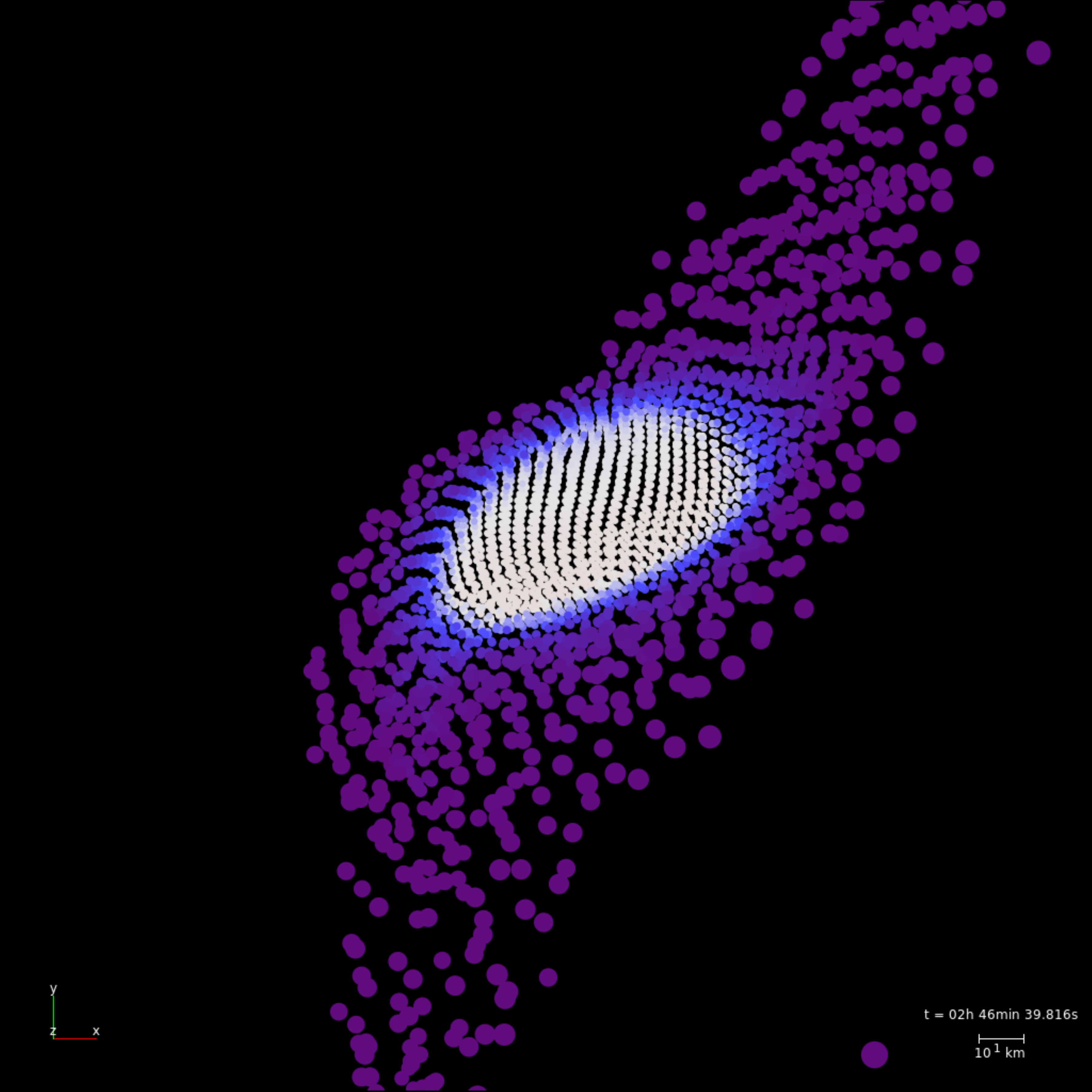} &
\includegraphics[width=4.6cm]{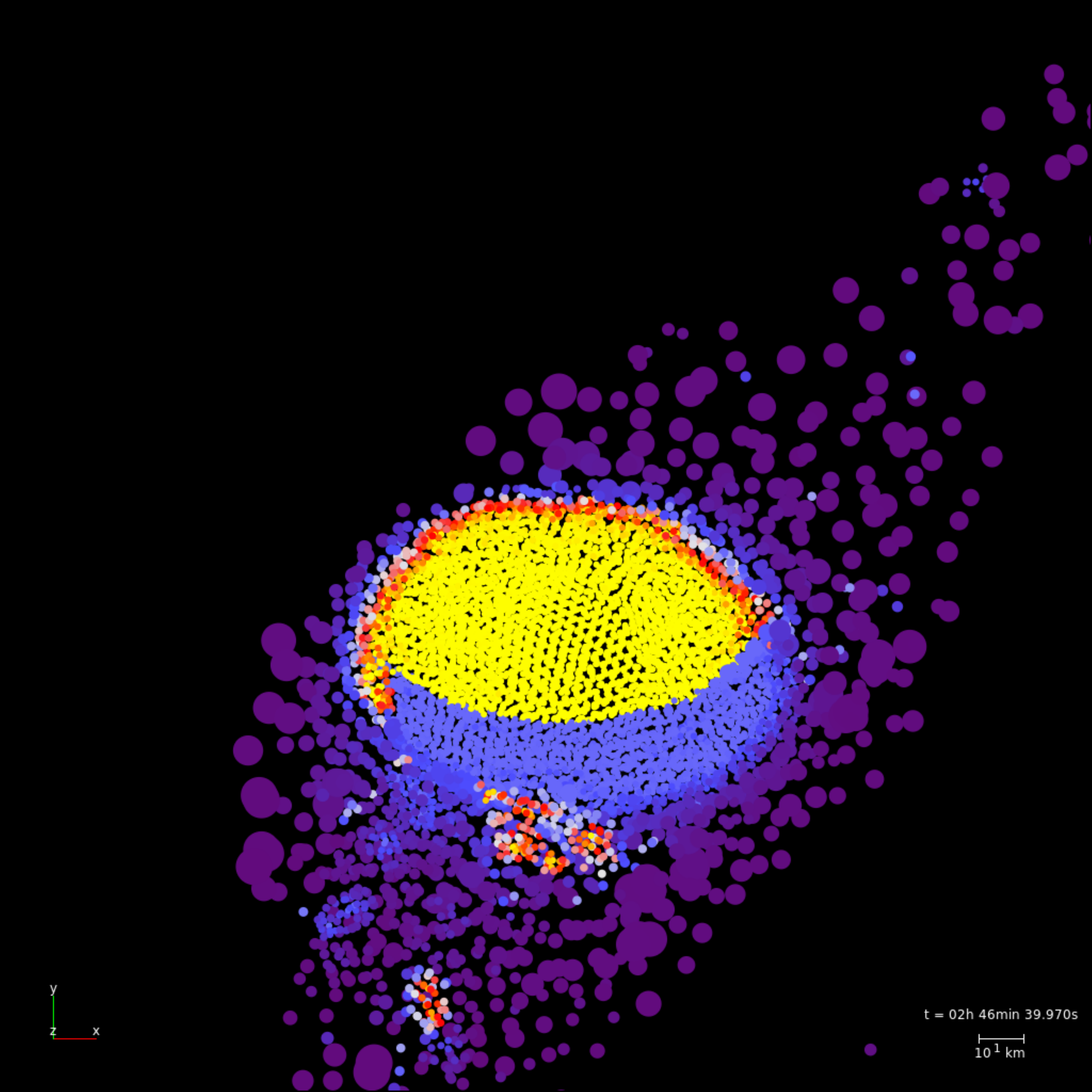} \\[.1cm]
\includegraphics[width=4.6cm]{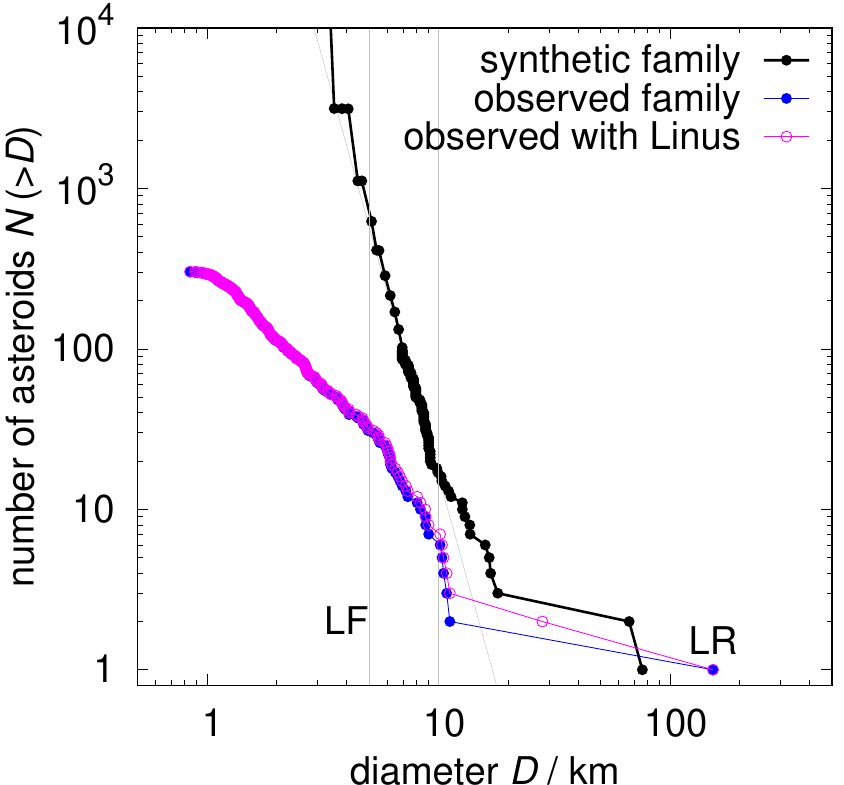} &
\includegraphics[width=4.6cm]{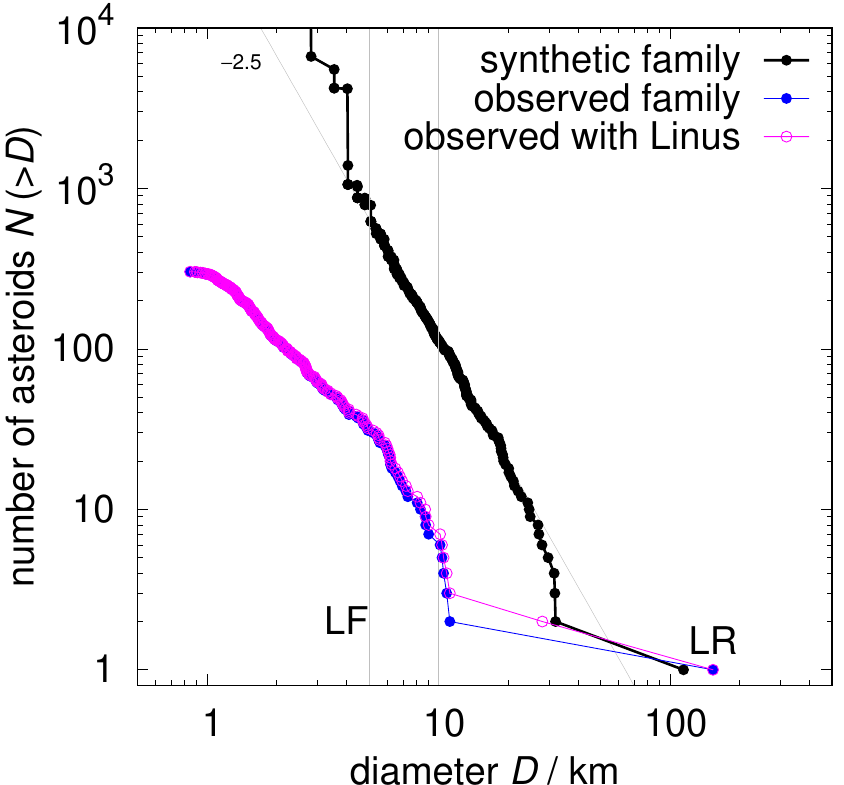} \\[.1cm]
OVERSHOOT &
WITH LINUS \\
\end{tabular}
\caption{(cont.)}
\label{kalliope10_homo_Maclaurin_165_45_6_0_size_distribution}
\end{figure*}

We solved the hydrodynamic equations for the three phases:
stabilisation (100\,s),
fragmentation (10\,000\,s),
reaccumulation (300\,000\,s).
In the first and second phases, we used
an asymmetric SPH solver,
adaptive smoothing lengths,
but the maximum smoothing length~$h$ was set to 17\,500\,m,
or 5.1 of the initial~$h$,
to prevent excessive cpu time due to expanded particles at the core-mantle boundary;
the rotation correction tensor,
a hash map for nearest-neighbors,
artificial viscosity,
self-gravity,
the opening angle~0.5, and
the multipole order~3
\citep{Sevecek_2021PhDT........36S}.
In the third phase, we started with
an equal-volume hand-off, and continued with
a simplified N-body solver,
merge-or-bounce collision handler,
repel-or-merge overlap handler,
the normal restitution~0.5,
the tangential restitution~1,
the merge velocity limit~4 (see explanation in \citealt{Sevecek_2021PhDT........36S}, p.~120), and
the merge rotation limit~1.
The total number of particles was~$10^5$.

Before we proceed, let us recall basic parameters of the Earth-Moon-forming impact
for comparison. According to \citep{Canup_2014RSPTA.37230175C}, it occurred with
the impact speed $v_{\rm imp}/v_{\rm esc} < 1.1$,
the impact parameter $b \simeq 0.7$,
and the projectile mass fraction $\gamma \simeq 0.125$.
The Moon accreted from a circumplanetary disk.
In this scenario, the disk mass is high if $v_{\rm imp}\downarrow$, $b\uparrow$, $\gamma\uparrow$;
the disk mass is low if $v_{\rm imp}\uparrow$;
and it is variable if $b < 0.7$.

In the case of (22)~Kalliope, the relative speed, when normalized by $v_{\rm esc}$,
is orders of magnitude higher,
$v_{\rm imp}/v_{\rm esc} \simeq 50$,
i.e., a totally different regime.
The projectile angular momentum (AM) for an intermediate angle
was of the order of
$L_{\rm imp} \doteq 2\times 10^{25}\,{\rm kg}\,{\rm m}^2\,{\rm s}^{-1} \doteq 1.9\,L_{\rm rot}$,
i.e., comparable to the {\rm current} rotational angular momentum of (22)~Kalliope.
However, in the reaccumulative regime \citep{Vernazza_2020NatAs...4..136V} one can hardly expect AM embedding;
AM draining is more likely \citep{Sevecek_2019A&A...629A.122S},
when the impact ejects material from an already rotating target.

Our results are summarized in Fig.~\ref{kalliope4_homo_DP_153_29_6_45_size_distribution}.
The impacts actually span a range of energies,
from low-energy, which did not eject enough fragments (i.e., `undershoot' the SFD)
to high-energy, which ejected 10 times more (`overshoot').
This was needed to understand overall trends.
At the same time, we monitored a shape of the largest remnant (a.k.a. (22) Kalliope);
the respective changes were
from minor
to major,
which is to be expected in the reaccumulative regime.

\begin{figure}
\centering
\includegraphics[width=8.9cm]{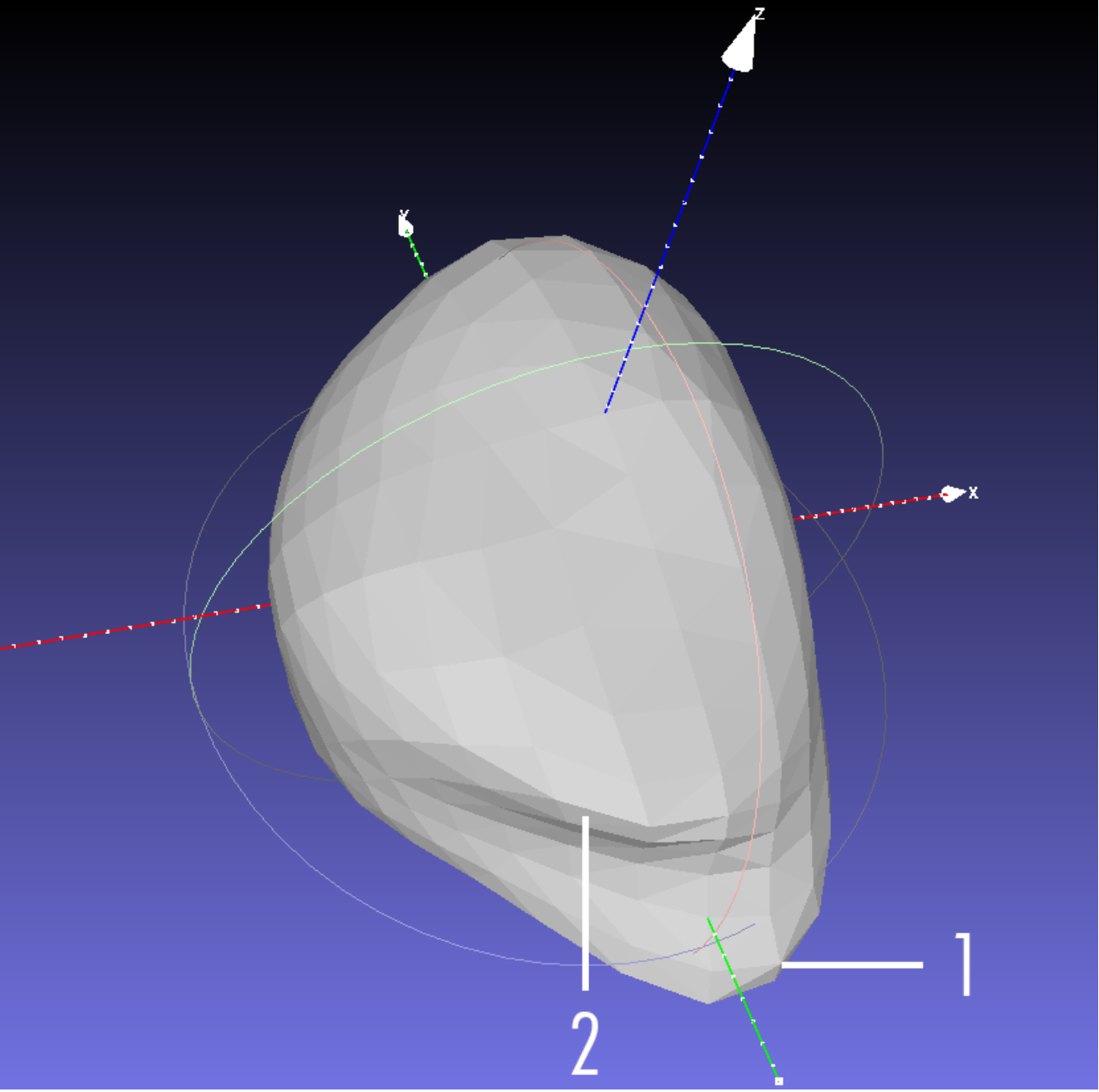}
\caption{Shape of (22) Kalliope according to the ADAM model,
with two `hills' approximately in $-\hat y$ direction,
separated by approximately $15^\circ$ in longitude.}
\label{kalliope_meshlab}
\end{figure}

\paragraph{Low-energy impacts.}
In the case of low-energy, spherical, $45^\circ$ impacts
(Fig.~\ref{kalliope4_homo_DP_153_29_6_45_size_distribution}, 1st, 2nd columns),
the SFD is undershoot;
the largest fragment (LF) is only 5\,km,
but the observed LF is 10\,km.
While this is clearly a poor fit,
it is important to note the shape of the largest remnant (LR).
We can monitor it till the end of the fragmentation phase;
it is determined by two general processes:
(i)~ejection of material from around the antipode,
which falls back to the surface a short time scale,
but {\em not\/} back at the antipode,
because the LR started to rotate due to the AM draining;
(ii)~excavation at the impact point,
with material flying on ballistic trajectories,
which falls back on the keplerian time
on the other hemisphere,
again rotated in the meantime.
This creates {\em two\/} characteristic `hills' on the surface
(see the animation associated to Fig.~\ref{kalliope4_homo_DP_153_29_6_45_size_distribution}).
Interestingly, it is very similar to the two hills in the $-\hat y$ direction,
observed at the surface of (22) Kalliope (Fig.~\ref{kalliope_meshlab}).
We thus cannot exclude the possibility that they were created late,
by a low-energy impact.
Because material is fully damaged, the coefficient of friction is crucial
for the obtained shape (as in \citealt{Vernazza_2020NatAs...4..136V}).

Another general feature is a flatter surface created by excavation,
perpendicular to the impact direction.
It is similar to the observed shape in $+\hat y$ direction.
The overall shape remains too close to spherical though;
at least medium-energy would be needed.

If the interior is differentiated, the surface is even flatter,
because the core is denser and its moment of inertia slows-down its motion.
Moreover, the core after impact is no longer spherical,
it is also flatter
and much closer to the surface on the side of impact.
This configuration is non-hydrostatic, though, and might further evolve over geologic time scales.

\paragraph{Medium-energy impacts.}
For medium-energy impacts into Maclaurin ellipsoids
(Fig.~\ref{kalliope4_homo_DP_153_29_6_45_size_distribution}, 3rd, 4th columns)
the outcome was variable.
For a homogeneous body,
the SFD is overshoot,
containing an intermediate-size fragment
(between LR and LF).
We believe it should be possible to find a better solution for the SFD,
but we did not find it in our limited set of simulations.
Nevertheless, the overall shape is now elongated enough,
with the two peaks still present.
From a broader perspective, reaccumulation has a form of `streams',
with material gravitationally attracted from larger distances.

For a differentiated body,
ejection is sufficient to match the LF, but not Linus.
Asymmetry of the core is even more pronounced.
Another difference is that the core is closer to the surface elsewhere,
as in the $+\hat x$ direction on Fig.~\ref{kalliope_meshlab}.
Even though we do not have enough resolution to see fine-grained ejecta,
one side of the core was fully exposed in the course of impact,
and we expect that metallic ejecta must partly cover the surface.
This is fully compatible with the M-type taxonomy of (22) Kalliope.

\paragraph{High-energy impacts.}
Increasing energy
(Fig.~\ref{kalliope4_homo_DP_153_29_6_45_size_distribution}, 5th, 6th columns)
further leads to overshooting the SFD.
Nevertheless, the highest energy actually produced Linus as the LF,
but it is on unbound orbit.
We also do not see any disk-like structure,
from which a bound moon could be formed.
The slope is steep from $D = 30\,{\rm km}$,
and the total number of fragments is 10 times larger than observed,
i.e., similar to the old `populous' family from Sec.~\ref{collisional}.
However, we find it difficult to eliminate 100 $D > 10\,{\rm km}$
bodies from the SFD by long-term evolution,
in order to match the observed SFD.
It would be in contradiction also with chaotic diffusion of (22)~Kalliope,
as discussed in Sec.~\ref{orbital}.

The LR is substantially smaller than the parent body.
For a homogeneous composition and a head-on impact,
it led to splitting and a pair of unbound similarly-sized LRs.
This is in contradiction with observations of Kalliope,
but it is a logical continuation of the trend from low- to medium- to high-energy impacts.
(A small-sized version of this phenomenon was studied by
\citealt{Vokrouhlicky_2021A&A...654A..75V}.)

For a differentiated body, the core is exposed even more,
because reaccumulation is not so efficient.
At the same time, one may expect late secondary impacts
of metallic material.
Again, this is compatible with the M~type.

\vskip\baselineskip

During break-ups, the final densities of the LR, the LF and other fragments
are generally different from the initial densities.
Ejection of low-density mantle material is a logical explanation
for the high density of (22) Kalliope.
At the same time, Linus --if created during this break-up--
should be composed of mantle material and have low density,
analogously to the Earth--Moon system.
Unfortunately, our model does not allow us to estimate the final densities
accurately for three reasons:
(i)~at the end of the fragmentation phase, material compression is still ongoing;
(ii)~during the reaccumulation phase, densities are fixed (by the hand-off);
(iii)~our model does not contain a treatment of porosity and compaction.
What we can do in the future is to prolong our computations $30$~times,
so that reaccumulation is also treated in the SPH framework.

Of course, we did not fully explore the parameter space, e.g., for
high initial rotation,
oblique,
non-equatorial,
or retrograde impacts.
Fast initial rotation might help to create a massive moon (Linus).
Nevertheless, we think the initial rotation was not too fast (close to critical),
because the two `hills' then would be too separated from each other.
Moreover, impacts should have some $\hat z$ velocity component,
because the hills have slightly different $z$ coordinates.

%%%%%%%%%%%%%%%%%%%%%%%%%%%%%%%%%%%%%%%%%%%%%%%%%%%%%%%%%%%%%%%%%%%%%%%%

%%%%%%%%%%%%%%%%%%%%%%%%%%%%%%%%%%%%%%%%%%%%%%%%%%%%%%%%%%%%%%%%%%%%%%%%

\section{Discussion}

\subsection{Stochasticity in SPH simulations.}

A late phase of gravitational reaccumulation,
even if it is described by a fully hydrodynamical (SPH) simulation,
corresponds to a few-$N$-body problem,
because there are only a few big bodies left.
Such systems are known to exhibit deterministic chaos
(e.g., \citealt{Nekhoroshev_1977RuMaS..32....1N}).
In our case, the very existence of Linus in our simulations,
and whether it is bound or unbound,
likely depends on few collisions,
which either occur or not.

The Earth-Moon-forming impacts also exhibit a broad range of outcomes
\citep{Canup_2019}
and only some of them are analogous to the Earth--Moon system.
Consequently, we expect more simulations would be needed (up to $10^2$)
to fit the Kalliope--Linus system (or complete SFD).
Yet, the number of free parameters is of the order of $10^1$,
to describe the respective
geometry,
rotation,
energetics,
as well as materials.
We thus postpone a computation of a corresponding `matrix' of simulations
for future work.

%%%%%%%%%%%%%%%%%%%%%%%%%%%%%%%%%%%%%%%%%%%%%%%%%%%%%%%%%%%%%%%%%%%%%%%%

\subsection{Constraining the origin of Kalliope}

\begin{figure}
\centering
\includegraphics[width=6.5cm]{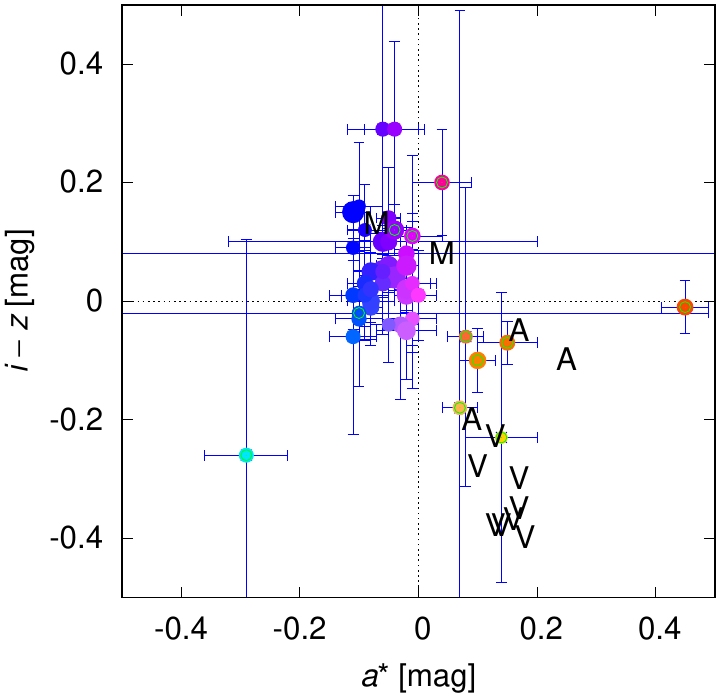}
\caption{SDSS colour indices $a^\star$ versus $i-z$
\citep{Parker_2008Icar..198..138P} for the Kalliope family.
Most of them correspond to C/X-complex taxonomy,
or to M-type if albedo is also taken into account.
A subset of bodies with known colours is plotted (circles with error bars);
probable interlopers are indicated by green circles.
Examples of taxonomic classes related to differentiated bodies
are also indicated by letters ('V', 'A', 'M');
these are not family members.}
\label{astar_iz}
\end{figure}

The Kalliope collisional family is the second known family related to a
differentiated body after that of Vesta. As such, it allows investigating and
characterizing the differentiation process on such early formed body. In our
current understanding, differentiated bodies comprise mostly V-types
(basaltic), A-types (olivine) and M-types (metallic). Both spectroscopic
observations of the primary and the Sloan Digital Sky Survey (SDSS;
\citealt{Parker_2008Icar..198..138P}) colors of the family members
(Tab.~\ref{tab:tax} and Fig.~\ref{astar_iz}) imply a C/X-type classification
for nearly all family members. When adding the albedo information for these
objects (these possess moderate optical albedos in the [0.1; 0.35] range), it
appears that essentially all family members are M-type asteroids alike (22)
Kalliope. In our search for possible A- and V-type family members, we
identified 6 bodies that exhibit colours similar to S-types (38309, 112382,
127063, 145265, 2002~OP$_{6}$, 373880), or possibly V-types in one case
(373880). Given that the orbital properties of most of these bodies are quite
different from those of the core of the family, we consider them as
interlopers.

The prevailing M-type classification confirms the absence of olivine
((Mg,Fe)$_2$SiO$_4$), this mineral being the standard one expected for mantles
of differentiated bodies. Actually, olivine-rich bodies are rare everywhere in
the asteroid belt \citep{DeMeo_2013Icar..226..723D, DeMeo_2019Icar..322...13D}
which led authors to suggest that the parent bodies of differentiated
meteorites may have been battered to bits \citep{Burbine_1996M&PS...31..607B}.
The formation of an olivine-rich mantle may, however, not have been the norm,
especially if bodies such as Kalliope formed beyond the snowline among the
later formed carbonaceous chondrite (CC)-like bodies. As suggested by
\citet{Hardersen_2005Icar..175..141H}, if Kalliope's parent body initially
contained as much carbon as found in CCs along with iron-bearing olivine as
commonly found in CO, CV or CR chondrites, then it could have experienced a
smelting-type reaction provided that it experienced internal temperature above
$850\,^\circ{\rm C}$. In Kalliope's case, such high internal temperature is
expected given its differentiated interior and metal-rich core. The
smelting-type reaction implies that if sufficient carbon is present as a
reducing agent, the final products would be enstatite (MgSiO$_3$), or other
iron-poor pyroxene, metallic iron, and possibly silica (SiO$_2$)
\citep{Hardersen_2005Icar..175..141H}.

A formation beyond the snowline for a large fraction of main-belt M-type
asteroids would be consistent with the majority of these bodies residing in the
outer belt \citep{DeMeo_2013Icar..226..723D}, a region essentially populated by
bodies (C, P and D-types) having likely formed beyond the snowline
\citep{Vernazza_2021A&A...654A..56V}. (22)~Kalliope may as such be a likely
parent body of CC related iron meteorites \citep{Kruijer_2017PNAS..114.6712K}.

%%%%%%%%%%%%%%%%%%%%%%%%%%%%%%%%%%%%%%%%%%%%%%%%%%%%%%%%%%%%%%%%%%%%%%%%

\section{Conclusions}

In this work, we studied the Kalliope family.
First, it was difficult to find,
because it has been dispersed, and
because the orbit of (22) Kalliope has been changed by chaotic diffusion.
As we already know, the hierarchical clustering method has limitations,
for example,
to identify family halos \citep{Broz_2013Icar..223..844B},
to recognize interlopers, or
it may fail to identify the largest remnant,
(22)~Kalliope.
Families may be even harder to find than previously thought,
especially if they are old, dispersed and depleted by long-term orbital and collisional evolution.
Nonetheless, satellites (smashed-targets, or SMATS, in particular) are good indicators of families.
It independently confirms that (107)~Camilla and (121)~Hermione,
with non-existent families but existing satellites,
may have had families \citep{Vokrouhlicky_2010AJ....139.2148V},
prior to early instabilities in the Solar System.

Second, according to our simulations, it is possible to outline
a consistent scenario.
The parent body broke-up $(900\pm100)\,{\rm My}$ ago,
due to an impact with $29$ to $45\,{\rm km}$ projectile.
(22)~Kalliope was partly preserved and partly reaccumulated;
ejected material was mostly from the silicate mantle,
which explains why (22)~Kalliope,
with preserved iron core,
has the exceptional density $4.1\,{\rm g}\,{\rm cm}^{-3}$.
We successfully explain formation of hills on the surface by reaccumulation.
Unfortunately, we did not find a simulation in our limited set,
which would produce a 1:1 counterpart of the observed moon Linus,
on a bound orbit, but we have `cornered' a parameter space,
in which a solution shall be found.
Long-term collisional as well as orbital evolution then explains
not only the very shallow size-frequency distribution,
but also uneven spread in eccentricity due to the 17:7 resonance,
offset of (22) Kalliope in eccentricity due to the $4{-}1{-}1$ resonance,
all of them on the correct time scale.
(See also Appendix~\ref{linus} for a~discussion of Linus orbit.)

Third, if the parent body was differentiated, axially symmetric,
with the centre of mass coinciding with the centre of volume,
after break-up the asymmetry of the internal structure is so substantial,
that it should also affect the dynamics of the moon orbit.
We predict that the iron core is close to the surface
on one side of the body,
likely the one which is flatter (in $+\hat y$ direction; Fig.~\ref{kalliope_meshlab}).
The surface of this M-type asteroid is unevenly covered
with metallic ejecta.
All these features should be eventually constrained by observations,
e.g., by spatially-resolved polarimetric measurements, or
reflex motion with respect to a suitable reference.

Given systematic uncertainties of densities of other fragments
(family members), it would be very useful to systematically
search for binaries among them.
If there are 1/6 of binaries in the main belt,
and presumably old family is evolved similarly as the main belt,
we expect 50 binaries,
either escaping-ejecta (EEB) or YORP-spin-up.
Some of them should exhibit eclipses,
which is an opportunity to determine the average density of components as:
\begin{equation}
\rho = {3\pi\over G}\left[\left({R_1\over a}\right)^3\!\!+\left({R_2\over a}\right)^3\right]^{-1}\,,
\end{equation}
where
$R_1$, $R_2$~denote their radii and $a$~the semimajor axis.
If the value will be lower than Kalliope's average density,
it will be an independent confirmation of its differentiation,
because we expect them to originate from Kalliope's mantle.
It may be difficult if not impossible to distinguish it from
alternative processes thought --- in particular,
ejected material is expanded during the fragmentation phase,
reaccumulated as a porous material,
and compacted on possibly long time scales.

% ??? PB->LR change of bulk density in SPH; long-term simulation?
% ??? compute SFD's correctly for differentiated bodies (material ID`)!

%%%%%%%%%%%%%%%%%%%%%%%%%%%%%%%%%%%%%%%%%%%%%%%%%%%%%%%%%%%%%%%%%%%%%%%%

\begin{acknowledgements}
This work has been supported by the Czech Science Foundation through grant
21-11058S (M.~Bro\v z).
We thank the referee A.~Morbidelli for comments,
which helped us to re-think implications of our work.
\end{acknowledgements}

\bibliographystyle{aa}
\bibliography{references}

%%%%%%%%%%%%%%%%%%%%%%%%%%%%%%%%%%%%%%%%%%%%%%%%%%%%%%%%%%%%%%%%%%%%%%%%

\cleardoublepage
\appendix

\section{Evolution of Linus orbit}\label{linus}

We also estimated the time scale of evolution of Linus orbit.
For the tidal torque acting on Linus,
we applied the standard formula \citep{dePater_2010plsc.book.....D}:
\begin{equation}
{\Gamma\over L} \simeq {3\over 2}{k_2\over Q}{Gm_2^2 R_1^5\over a^6} \left({m_1m_2\over m_1+m_2}\sqrt{G(m_1+m_2) a}\right)^{-1} \,,\label{Gamma_L_tidal}
\end{equation}
where
$\Gamma$~denotes the torque,
$L$~orbital angular momentum,
$a$~semimajor axis.
$m_2$~mass of the perturbing body (Linus),
$R_1$~radius of the perturbed body (Kalliope),
$m_1$~its mass,
$k_2$~the Love number, and
$Q$~the dissipation factor.
The free parameter is the ratio $k_2/Q$.
Unfortunately, it is unconstrained by observations.
According to our tests, a model with tides
\citep{Mignard_1979M&P....20..301M,Broz_2022A&A...657A..76B}
is statistically equal to a model without tides.
Nevertheless, we can assume tides either strong
($Q = 40$, $k = 0.305$ as for (216) Kleopatra; \citealt{Broz_2022A&A...657A..76B}),
or weak
($Q = 280$ as for the Earth, $k = 0.024$ as for Moon).

Similarly, for the radiative torque on synchronous satellites
we applied the scaling from \cite{Cuk_2005Icar..176..418C}:
\begin{equation}
{|\Gamma|\over L} \simeq 3.0\cdot10^{-12}\,{\rm s}^{-1}\,\left({a_{\rm h}\over a_{\rm h0}}\right)^{-2} \left({\rho\over\rho_0}\right)^{-1} \left({a_1\over a_{10}}\right)^{-1} \left({R_2\over R_{20}}\right)^{-1} {P_1\over P_{10}}\,,\label{Gamma_L_radiative}
\end{equation}
where
$a_{\rm h}$ denotes the heliocentric semimajor axis,
$\rho$ density,
$a_1$ the binary semimajor axis,
$R_2$ the radius of the secondary,
$P_1$ the binary period.
The quantities with $0$ subscripts are normalisations.
Again, the radiative effects can be strong
(the coefficient as above),
or weak
($0.5\cdot 10^{-12}\,{\rm s}^{-1}$).
The sign of $\Gamma$ is either positive or negative;
it depends on the detailed (unknown) shape of the moon.

\begin{figure}
\centering
\includegraphics[width=8.5cm]{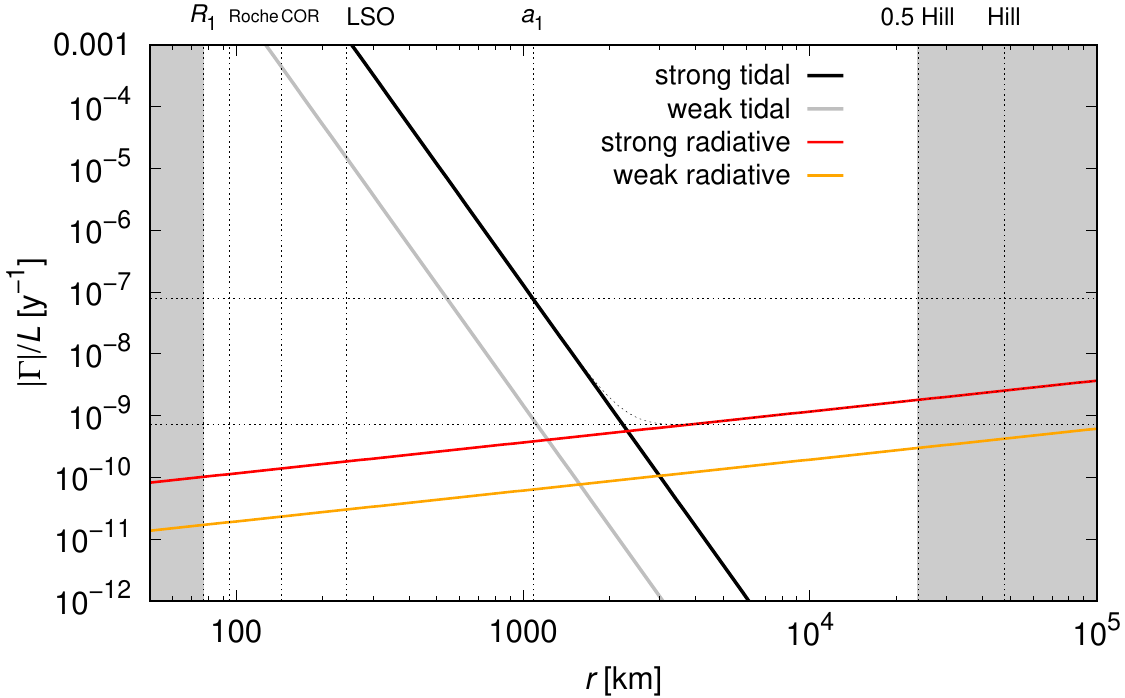}
\caption{Expected torque per angular momentum $|\Gamma|/L$
versus distance~$r$ in the Kalliope--Linus system.
We separately plot the tidal (black, gray) and the radiative (red, orange) torques.
We also distinguish two levels of the tidal dissipation
as well as two levels of the radiative effects.
Linus is located at $a_1 \doteq 1060\,{\rm km}$,
i.e. close to a possible equilibrium between the torques.}
\label{vypocty20_BYORP}
\end{figure}

The comparison (Fig.~\ref{vypocty20_BYORP}) shows that
the two torques should be comparable between $1200$ and $3000\,{\rm km}$.
Interestingly, Linus is located at $1060\,{\rm km}$,
very close to an equilibrium between the weak positive tidal and
strong negative radiative torques.
It seems reasonable that tidal dissipation is weaker in (22) Kalliope
than in (216) Kleopatra, because the former is likely differentiated
(more rigid, less viscous) and certainly not-so-extreme as the latter.

The time scale of evolution, starting from the Roche radius,
corotation orbit (COR), or the last stable orbit (LSO),
is about $10^8\,{\rm y}$ up to the location of Linus,
but of course an approach to the exact equilibrium
is very slow ($5\cdot 10^8\,{\rm y}$).
If both torques were positive and there is no equilibrium,
the overall evolution up to the 0.5 Hill radius would take $2\cdot 10^9\,{\rm y}$.
Clearly, even weak tides are sufficient to explain the evolution of Linus
well within the minimum dynamical age of the Kalliope family.

%%%%%%%%%%%%%%%%%%%%%%%%%%%%%%%%%%%%%%%%%%%%%%%%%%%%%%%%%%%%%%%%%%%%%%%%

\section{List of family members}

We list all 302 family members (with interlopers removed):

\begingroup
%\raggedright
\footnotesize
\vskip\baselineskip
\noindent
(22)~Kalliope,
(22)~Linus,
  7481,
 12573,
 14012,
 14338,
 16367,
 17845,
 17994,
 18733,
 21216,
 21741,
 22087,
 22108,
 24879,
 25631,
 26023,
 26213,
 27916,
 28089,
 31064,
 31709,
 32646,
 34565,
 37292,
 37669,
 41358,
 41930,
 46212,
 46272,
 47061,
 47898,
 50806,
 51259,
 53350,
 54280,
 54908,
 55418,
 62032,
 68222,
 71357,
 71396,
 73446,
 82711,
 82731,
 82787,
 85417,
 94038,
 94199,
 95912,
104173,
110549,
111144,
111199,
112430,
112631,
114664,
116170,
119834,
131265,
140031,
140108,
141200,
150541,
152341,
152606,
153086,
154059,
159032,
161396,
161565,
161837,
164369,
166751,
171194,
175716,
177128,
178329,
180960,
181005,
190114,
196897,
196974,
203800,
206688,
216530,
217802,
226746,
227719,
228284,
229227,
232543,
237969,
245390,
246484,
248020,
250675,
254820,
254988,
261583,
267786,
277936,
279504,
280659,
284640,
285991,
286015,
295381,
296463,
296943,
304544,
305259,
306388,
316440,
318919,
319447,
322682,
325905,
326039,
327530,
328400,
329280,
331508,
331810,
333674,
334501,
334874,
338276,
351859,
357736,
360902,
361395,
363398,
366347,
368462,
368809,
378902,
380475,
387687,
396582,
400928,
403919,
404857,
409327,
410858,
413399,
414629,
415732,
420228,
423037,
426532,
427317,
428933,
430340,
440414,
444625,
454161,
454418,
462450,
464107,
464304,
464612,
472463,
477110,
483146,
485768,
487450,
488123,
492163,
501370,
503314,
505186,
505266,
507356,
510489,
515640,
518825,
520740,
521968,
523413,
526488,
529676,
532317,
532409,
533253,
536305,
538932,
539027,
539263,
545524,
546186,
549084,
551844,
555164,
559437,
560485,
562647,
562836,
563578,
563959,
564493,
566962,
569114,
569157,
571468,
573702,
576659,
576709,
576969,
581697,
582004,
582022,
583589,
583772,
588322,
590269,
590279,
594884,
598037,
600315,
601088,
601468,
604253,
1999~FS$_{99}$,
2000~SX$_{379}$,
2002~CB$_{315}$,
2002~RC$_{265}$,
2003~SE$_{228}$,
2004~KB$_{20}$,
2005~TY$_{83}$,
2005~VY$_{137}$,
2005~WN$_{116}$,
2006~CW$_{31}$,
2006~YF$_{37}$,
2007~EQ$_{145}$,
2007~FY$_{54}$,
2007~RU$_{52}$,
2007~TX$_{459}$,
2007~VN$_{347}$,
2007~VX$_{305}$,
2008~EW$_{42}$,
2008~JF$_{12}$,
2008~KB$_{45}$,
2008~SM$_{319}$,
2008~TC$_{197}$,
2008~UH$_{232}$,
2009~SA$_{113}$,
2010~FU$_{140}$,
2010~RS$_{77}$,
2010~TT$_{192}$,
2010~YY$_{3}$,
2011~CW$_{125}$,
2011~EX$_{91}$,
2011~FR$_{66}$,
2011~UO$_{420}$,
2012~BG$_{62}$,
2012~GG$_{42}$,
2012~UY$_{194}$,
2012~XM$_{102}$,
2013~HS$_{125}$,
2013~VA$_{36}$,
2014~EG$_{244}$,
2014~HH$_{165}$,
2014~MO$_{66}$,
2014~NZ$_{22}$,
2014~QH$_{248}$,
2014~QN$_{72}$,
2014~QV$_{508}$,
2014~QX$_{473}$,
2014~QZ$_{103}$,
2014~RL$_{31}$,
2014~SV$_{118}$,
2014~WO$_{214}$,
2014~XX$_{3}$,
2015~BR$_{379}$,
2015~DH$_{253}$,
2015~DO$_{258}$,
2015~FH$_{411}$,
2015~GZ$_{21}$,
2015~HG$_{4}$,
2015~KB$_{83}$,
2015~KD$_{83}$,
2015~MP$_{57}$,
2015~MZ$_{90}$,
2015~SQ$_{24}$,
2015~TH$_{272}$,
2015~UH$_{30}$,
2015~VU$_{85}$,
2015~VZ$_{37}$,
2015~XB$_{413}$,
2015~XG$_{358}$,
2015~XQ$_{277}$,
2016~AO$_{141}$,
2016~CP$_{87}$,
2016~EL$_{258}$,
2016~KT$_{10}$,
2016~QV$_{66}$,
2016~UF$_{8}$,
2017~BD$_{122}$,
2017~SM$_{30}$,
2017~UJ$_{49}$,
2018~LS$_{17}$,
2018~OK$_{1}$.

\endgroup

% Table kalliope familly members
%\input{alb_tax}

\onecolumn
\begin{longtable}{llllll}
\caption{\label{tab:tax}Compilation of known albedos and taxonomic types of the Kalliope family members.}\\
\hline
Number &             Name &      Designation & $p_V$ & $\sigma$ of $p_V$ & Taxonomy \\
\hline\hline
    \endfirsthead

    \caption{continued}\\ 
    \hline
Number &             Name &      Designation & $p_V$ & $\sigma$ of $p_V$ & Taxonomy \\
    \hline\hline
    \endhead

    \hline
    \endfoot
    \hline
    \endlastfoot 
    22 &         Kalliope &    A852 WA &  0.166 &      0.005 &       - \\
  7481 &     San Marcello &   1994 PA1 &   0.17 &      0.074 &       M \\
 14012 &           Amedee &    1993 XG &  0.201 &      0.018 &       - \\
 14338 &      Shibakoukan &   1982 VP3 &  0.253 &       0.03 &       M \\
 16367 & Astronomiasvecia &   1980 FS4 &  0.247 &      0.045 &       - \\
 17845 &                - & 1998 HY112 &   0.13 &      0.021 &       - \\
 17994 &                - &  1999 JF70 &  0.199 &      0.016 &       - \\
 21216 &                - &   1994 UZ1 &   0.16 &      0.014 &       M \\
 21741 &                - & 1999 RN162 &  0.319 &      0.039 &       - \\
 22108 &                - &    2000 PD &  0.232 &      0.048 &       M \\
 24879 &                - &   1996 KO5 &      - &          - &     C/X \\
 25631 &                - &  2000 AJ55 &    0.2 &      0.039 &       M \\
 26023 &                - &   4538 P-L &   0.13 &      0.034 &       - \\
 26213 &                - &   1997 UV8 &   0.11 &      0.013 &       - \\
 28089 &                - &  1998 RD17 &   0.19 &      0.024 &       - \\
 31709 &                - &  1999 JD51 &  0.153 &      0.023 &       - \\
 32646 &                - &   3010 P-L &  0.234 &      0.033 &       M \\
 34565 &                - & 2000 SY292 &  0.138 &      0.029 &       - \\
 37292 &                - &  2001 AN34 &  0.172 &      0.029 &       - \\
 37669 &                - &   1994 TH1 &      - &          - &       U/X \\
 41358 &                - &  2000 AJ54 &  0.174 &      0.025 &       - \\
 46212 &                - & 2001 FD162 &  0.275 &      0.051 &       - \\
 46272 &                - &  2001 HO64 &  0.243 &      0.132 &       - \\
 47061 &                - &  1998 XZ43 &  0.195 &      0.031 &       - \\
 47898 &                - &  2000 GA47 &  0.105 &      0.008 &       - \\
 50806 &                - &  2000 FH28 &  0.163 &      0.009 &       M \\
 51259 &                - &  2000 JY59 &      - &          - &       C \\
 53350 &                - &  1999 JD65 &  0.196 &      0.064 &       M \\
 54280 &                - &  2000 JF47 &   0.35 &      0.077 &       - \\
 54908 &                - &  2001 OY80 &   0.18 &      0.025 &       - \\
 68222 &                - &  2001 CQ47 &  0.265 &      0.209 &       - \\
 71357 &                - & 2000 AJ122 &      - &          - &    C/CX \\
 71396 &                - & 2000 AV166 &    0.2 &       0.09 &       - \\
 73446 &                - &  2002 NX12 &  0.146 &      0.032 &       - \\
 82787 &                - &  2001 QP22 &      - &          - &       X \\
 85417 &                - &   1996 XQ3 &  0.225 &      0.083 &       - \\
 94199 &                - &  2001 BM16 &  0.173 &      0.042 &       - \\
104173 &                - &  2000 EE83 &  0.199 &      0.039 &       - \\
110549 &                - & 2001 TC101 &  0.167 &      0.048 &       - \\
111144 &                - &  2001 VH99 &  0.197 &      0.048 &       M \\
111199 &                - &  2001 WW21 &  0.127 &      0.012 &       - \\
112430 &                - &  2002 NJ51 &      - &          - &      XD \\
112631 &                - &  2002 PT77 &  0.261 &      0.244 &       M \\
119834 &                - &   2002 CK3 &      - &          - &       C \\
131265 &                - &  2001 FD43 &  0.103 &       0.03 &       - \\
141200 &                - & 2001 XP203 &  0.181 &      0.044 &       - \\
159032 &                - &  2004 TK67 &  0.224 &      0.043 &       M \\
161837 &                - &  2006 XZ63 &  0.246 &        0.1 &       - \\
166751 &                - &   2002 UZ2 &  0.185 &      0.039 &       M \\
178329 &                - &  1995 SO36 &      - &          - &     C/X \\
196974 &                - &  2003 UC64 &      - &          - &       C \\
227719 &                - & 2006 DK197 &      - &          - &       C \\
229227 &                - &  2004 XE18 &  0.273 &      0.083 &       - \\
254988 &                - & 2005 SY266 &      - &          - &      CX \\
285991 &                - &  2001 SH15 &      - &          - &      XD \\
322682 &                - &  1999 VY53 &  0.142 &      0.047 &       - \\
331508 &                - & 1999 XC261 &      - &          - &       C \\
338276 &                - & 2002 TZ309 &      - &          - &       C \\
387687 &                - & 2002 TW309 &      - &          - &       M \\
396582 &                - &  2000 RM78 &      - &          - &    C/CX \\
427317 &                - & 2014 WH292 &  0.101 &        0.1 &       - \\
     - &                - & 2014 QX473 &  0.067 &      0.022 &       - \\
\end{longtable}
\tablefoot{Albedo values are from the NEOWISE dataset
\citep{2011ApJ...741...68M,2012ApJ...759L...8M} and taxonomic types are
SDSS-based classification by\\ \citet{2010A&A...510A..43C} or
\citet{2013Icar..226..723D}. If the albedo is higher than 0.1, C/X-types were
re-classified as M.}

\end{document}